\documentclass[]{aastex631}

\shorttitle{PSP First Observation of the Two ICMEs interaction}
\shortauthors{Nieves-Chinchilla et al.}
\graphicspath{{./}{figures/}}

\hypersetup{linkcolor=blue,citecolor=blue,filecolor=blue,urlcolor=blue}

\begin{document}

\title{Direct First PSP Observation of the Interaction of Two Successive Interplanetary Coronal Mass Ejections in November 2020}

\correspondingauthor{Teresa Nieves-Chinchilla}
\email{teresa.nieves@nasa.gov}

\author[0000-0003-0565-4890]{Teresa Nieves-Chinchilla}
\affiliation{Heliophysics Science Division, NASA Goddard Space Flight Center, Greenbelt, MD 20771, USA}

\author[0000-0001-5207-9628]{Nathalia Alzate}
\affiliation{Heliophysics Science Division, NASA Goddard Space Flight Center, Greenbelt, MD 20771, USA}
\affiliation{ADNET Systems Inc., Greenbelt, MD 20771, USA}

\author[0000-0001-7080-2664]{Hebe Cremades}
\affiliation{Universidad Tecnol\'ogica Nacional, Facultad Regional Mendoza, CONICET, CEDS, M5505AJE Mendoza, Argentina}

\author[0000-0003-2361-5510]{Laura Rodríguez-García}
\affiliation{Universidad de Alcalá, Space Research Group, Alcalá de Henares, 28801 Madrid, Spain}

\author[0000-0001-5190-442X]{Luiz F. G. Dos Santos}
\affiliation{CIRES, University of Colorado at Boulder, Boulder, CO 80309, USA}


\author[0000-0001-6746-7455]{Ayris Narock}
\affiliation{Heliophysics Science Division, NASA Goddard Space Flight Center, Greenbelt, MD 20771, USA}
\affiliation{ADNET Systems Inc., Greenbelt, MD 20771, USA}

\author[0000-0002-0058-1162]{Hong Xie}
\affiliation{Department of Physics, The Catholic University of America, Washington, DC 20064, USA}
\affiliation{Heliophysics Science Division, NASA Goddard Space Flight Center, Greenbelt, MD 20771, USA}

\author[0000-0003-3255-9071]{Adam Szabo}
\affiliation{Heliophysics Science Division, NASA Goddard Space Flight Center, Greenbelt, MD 20771, USA}

\author[0000-0001-6590-3479]{Erika Palmerio}
\affiliation{Space Sciences Laboratory, University of California--Berkeley, Berkeley, CA 94720, USA}
\affiliation{CPAESS, University Corporation for Atmospheric Research, Boulder, CO 80301, USA}

\author[0000-0001-6185-3945]{Vratislav Krupar}
\affiliation{Goddard Planetary Heliophysics Institute, University of Maryland, Baltimore, MD 21250, USA}
\affiliation{Heliophysics Science Division, NASA Goddard Space Flight Center, Greenbelt, MD 20771, USA}

\author[0000-0002-1573-7457]{Marc Pulupa}
\affiliation{Space Sciences Laboratory, University of California--Berkeley, Berkeley, CA 94720, USA}

\author[0000-0002-3176-8704]{David Lario}
\affiliation{Heliophysics Science Division, NASA Goddard Space Flight Center, Greenbelt, MD 20771, USA}

\author[0000-0002-7728-0085]{Michael L. Stevens}
\affiliation{Harvard--Smithsonian Center for Astrophysics, Cambridge, MA 02138, USA}

\author[0000-0002-4313-1970]{Lynn B. Wilson III}
\affiliation{Heliophysics Science Division, NASA Goddard Space Flight Center, Greenbelt, MD 20771, USA}

\author[0000-0002-2106-9168]{Ryun-Young Kwon}
\affiliation{Korea Astronomy and Space Science Institute, Daejeon 34055, Republic of Korea}

\author[0000-0001-9177-8405]{M. Leila Mays}
\affiliation{Heliophysics Science Division, NASA Goddard Space Flight Center, Greenbelt, MD 20771, USA}

\author[0000-0001-8906-6097]{O. Chris St.~Cyr}
\affiliation{Heliophysics Science Division, NASA Goddard Space Flight Center, Greenbelt, MD 20771, USA}

\author[0000-0003-1377-6353]{Phillip Hess}
\affiliation{U.S. Naval Research Laboratory, Washington, DC 20375, USA}

\author[0000-0002-6903-6832]{Katharine K. Reeves}
\affiliation{Harvard--Smithsonian Center for Astrophysics, Cambridge, MA 02138, USA}

\author[0000-0002-0494-2025]{Daniel B. Seaton}
\affiliation{CIRES, University of Colorado at Boulder, Boulder, CO 80309, USA}
\affiliation{NCEI, National Oceanic and Atmospheric Administration, Boulder, CO 80305, USA}

\author[0000-0001-6692-9187]{Tatiana Niembro}
\affiliation{Harvard--Smithsonian Center for Astrophysics, Cambridge, MA 02138, USA}

\author[0000-0002-1989-3596]{Stuart D. Bale}
\affiliation{Space Sciences Laboratory, University of California--Berkeley, Berkeley, CA 94720, USA}
\affiliation{Physics Department, University of California--Berkeley, Berkeley, CA 94720, USA}

\author[0000-0002-7077-930X]{Justin C. Kasper}
\affiliation{BWX Technologies Inc., Washington, DC 20002, USA}
\affiliation{Department of Climate and Space Sciences and Engineering, University of Michigan, Ann Arbor, MI 48109, USA}



\begin{abstract}
We investigate the effects of the evolutionary processes in the internal magnetic structure of two interplanetary coronal mass ejections (ICMEs) detected in situ between 2020 November 29 and December 1 by Parker Solar Probe (PSP). The sources of the ICMEs were observed remotely at the Sun in EUV and subsequently tracked to their coronal counterparts in white light. This period is of particular interest to the community since it has been identified as the first widespread solar energetic particle event of Solar Cycle 25. The distribution of various solar and heliospheric-dedicated spacecraft throughout the inner heliosphere during PSP observations of these large-scale magnetic structures enables a comprehensive analysis of the internal evolution and topology of such structures. By assembling different models and techniques, we identify the signatures of interaction between the two consecutive ICMEs and the implications for their internal structure. We use multispacecraft observations in combination with a remote-sensing forward modeling technique, numerical propagation models, and in-situ reconstruction techniques. The outcome, from the full reconciliations, demonstrates that \textbf{the} two CMEs are interacting in the vicinity of PSP. Thus, we identify the in-situ observations based on the physical processes that are associated with the interaction and collision of both CMEs. We also expand the flux rope modeling and in-situ reconstruction technique to incorporate the aging and expansion effects in a distorted internal magnetic structure and explore the implications of both effects in the magnetic configuration of the ICMEs. 
\end{abstract}



\section{Introduction} \label{sec:intro}

It is well established that interplanetary coronal mass ejections (ICMEs) and their associated structures may unleash the most harmful geomagnetic storms at Earth and other undesired space weather effects \textbf{at spacecraft}. Unfortunately, little is known about their internal magnetic structure due to limitations imposed by currently-available measurements and techniques.  
The internal magnetic structure of coronal mass ejections (CMEs) is intrinsically associated with the initiation processes back at the Sun \citep[e.g.,][among others]{Mikic1994,Longcope2007,Cheng_2013,Patsourakosetal2013,Song2014, Patsourakos2020}. However, the CME transformation into an ICME is essentially due to physical processes related to its evolution and propagation \citep[see reviews by][]{Lugaz2017,Manchester2017,Luhmann2020}. How the internal magnetic structure of ICMEs \textbf{is} characterized or how long the innate CME features remain \textbf{untouched} in the interplanetary medium remain unanswered.

The progressive increase of ground-based assets, such as the K-Coronagraph (KCor), which is a part of the COronal Solar Magnetism Observatory \citep[COSMO;][]{Tomczyk2016}, and the space-based GOES-16/Solar Ultraviolet Imager \citep[SUVI,][]{Seaton2018,Vasudevan2019}, the Atmospheric Imaging Assembly \citep[AIA,][]{Lemen2012} instrument onboard the Solar Dynamics Observatory \citep[SDO;][]{Pesnell2012}, the Large Angle and Spectrometric Coronagraph \citep[LASCO;][]{Brueckner1995} onboard the Solar and Heliospheric Observatory \citep[SOHO;][]{Domingo1995}, and the Sun Earth Connection Coronal and Heliospheric Investigation \citep[SECCHI;][]{Howard2008} imaging suite onboard the Solar Terrestrial Relations Observatory \citep[STEREO;][]{Kaiser2008} provide a valuable combination of multiview remote-sensing observations that \textbf{enables} progress in the three-dimensional understanding of the physical processes associated with CME evolution.  With the advent of Parker Solar Probe \citep[PSP,][]{Fox2016} and Solar Orbiter \citep[SolO;][]{Muller2020}, multi-point in-situ analyses of large-scale structures in the solar wind has boosted the momentum created with MESSENGER \citep{Anderson2007} in the characterization of the internal magnetic structure of ICMEs \citep[][among others]{Nieves-Chinchilla2012,Nieves-Chinchilla2013,Good2015,Winslow2016,Davies2020,Lugaz2020}. 

\textbf{However, efforts to synchronize observations, researchers, resources, models and techniques towards reconciling the global and local views not only reveals the limitations in our understanding of the physical phenomena, but also the deficit of the techniques and models}.

There is still a scarcity of in-situ buoys in the interplanetary medium that prevents us from capturing the magnetic field and plasma \textbf{signatures} of ICMEs, which are needed to fully characterize their internal structure and track changes as they propagate and evolve through the heliosphere. As such, each multiview and multipoint event observed becomes a valuable opportunity to address these challenges. This is the case of the four successive CMEs that occurred during November 24--29 2020, listed in Table~\ref{tab:4CMEs}. The table lists the CME event name, time of first detection by the SOHO/LASCO C2 coronagraph, and central location at $\sim$20 solar radii. \textbf{This paper focuses on `CME1' and `CME2'.  However, `CME 01' and `CME 02' are considered as inputs for the WSA-ENLIL+Cone model \citep[hereafter ENLIL,][described in Section \ref{sec:Recons}]{Odstrcil2004}.  These CMEs can be important for the pre-conditioning of the interplanetary space, for example, lowering the density in which CME1 and CME2 propagate \citep[][]{Dumbovic2019}}.

CME2 gave rise to an unusual spread of energetic particles that are considered to be the first widespread solar energetic particle (SEP) event of solar cycle 25 \citep[][]{Cohen2021, Kollhoff2021, Lario2021, Kouloumvakos2021,Mitchell_2021}. Two out of the four CMEs observed by solar and heliospheric telescopes, namely CME1 and CME2, impacted PSP and left clear flux rope imprints detected by the magnetometer \citep[FIELDS;][]{Bale2016} onboard. \textbf{Maneuvers in the spacecraft during this period restricted the performance of the Solar Wind Electrons Alphas and Protons \citep[SWEAP;][]{Kasper2016} instrument, also onboard, and plasma information was therefore limited to the interplanetary counterpart of CME1 (ICME1)}. \textbf{Our work shows that the magnetic field signatures of ICME2, counterpart of CME2, are sufficient to identify the large-scale structure.  However, the lack of plasma information limits the full characterization of the internal magnetic field structure. To carry out the full in-situ characterization of ICME2, we will constrain the assumptions with remote-sensing observations, propagation models and in-situ observations of the flank of ICME2 at STEREO-A.  STEREO-A helps to characterize the event analysis in terms of the arrival time at STEREO-A and in-situ signatures.  In the case of ICME2, in-situ measurements do not display the magnetic obstacle (MO) magnetic field and plasma signatures}.

In this paper, we present an \textbf{extensive analysis of the successive ICME1 and ICME2 crossing the PSP spacecraft and the impact of evolutionary processes on these structures}. Section \ref{sec:observations} provides the event overview based on the observations available. 
Section \ref{sec:Recons} presents our analysis of the internal structure of the CMEs/ICMEs and a 3D reconstruction based on remote sensing and in-situ observations.  Given the lack of data on kinematic and plasma parameters, we have combined the available remote-sensing and in-situ observations with the ENLIL modeling simulation of the CMEs.  The reconciliation of the different techniques and models suggests that both \textbf{ICMEs} are interacting in the vicinity of PSP. Thus, Section \ref{Sec:discussion} provides insight into the in-situ observations that help evaluate the consequences of the interaction on the internal structure. As part of the discussion, we have implemented the effects of the aging and expansion in the \citet[][]{Nieves-Chinchilla2016,Nieves-Chinchilla2018ApJ} models in order to assess the magnetic configuration of an expanding and/or distorted magnetic flux rope. In Section \ref{Sec:conclusions} we summarize our results and conclusions.

\section{Events Overview} 
\label{sec:observations}

\begin{figure}[ht!]
    \centering
    \includegraphics[width=0.9\textwidth]{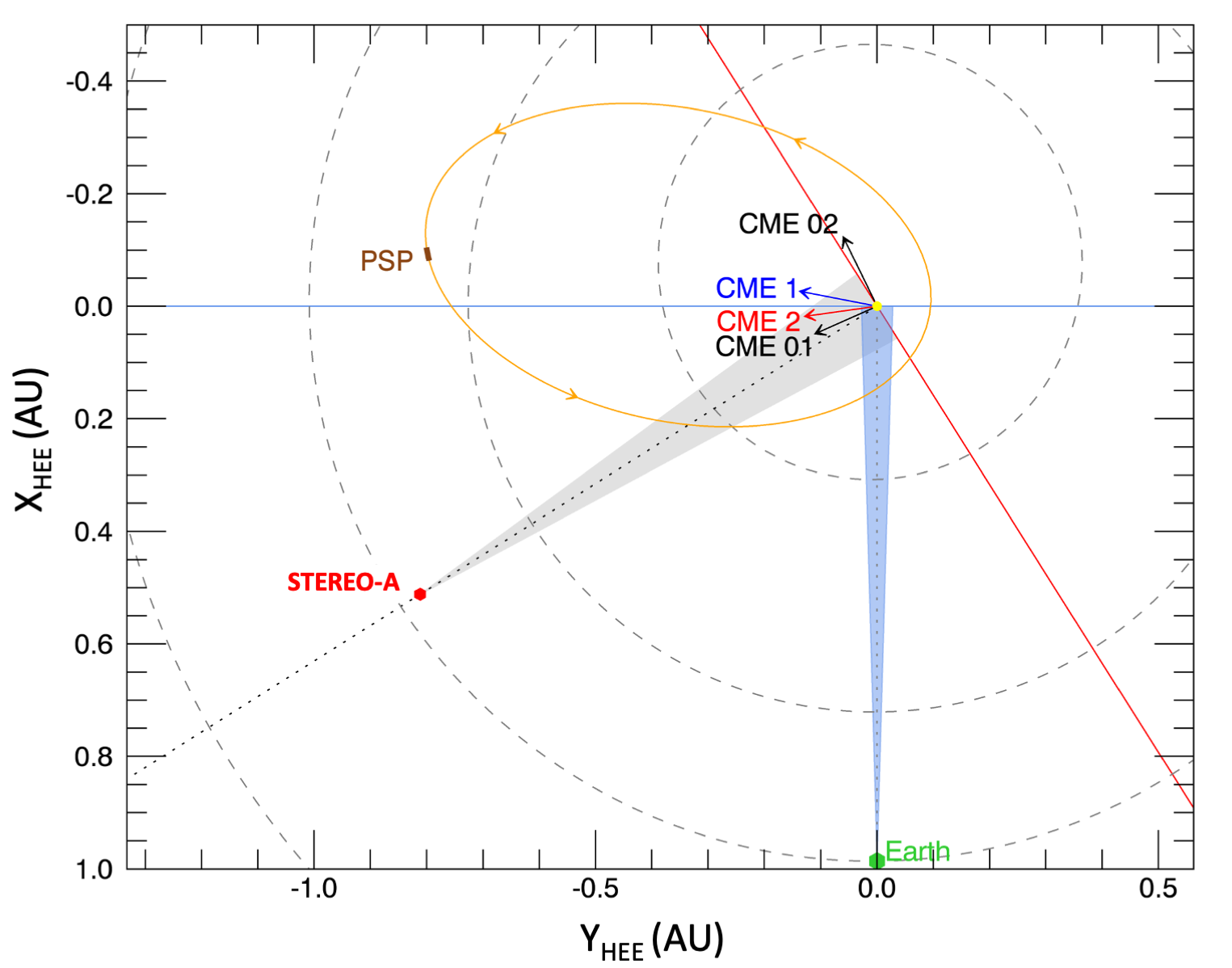}
    \caption{Map of the solar--heliospheric spacecraft location \textbf{(during the November 29 to December 1 time period)} and capabilities: PSP (orange), STEREO-A (A; red), and Earth (green) that includes SOHO, SDO, and GOES-16. The STEREO-A/COR2 field of view is delimited by the gray triangle and projected in the plane of the sky (red line) and the SOHO/C2 field of view is in blue color projected in the plane of the sky (blue line).The arrows represent the central direction of propagation of the CMEs listed in Table \ref{tab:4CMEs}.}
    \label{fig:spacecrafts}
\end{figure}

Figure~\ref{fig:spacecrafts} shows a map depicting the locations of multiple solar--heliospheric spacecraft \textbf{from 2020 November 29 to December 1 (during which PSP detected the transit of the ICMEs). PSP (orange) was moving from $\sim$96$^{\circ}$ to $\sim$97$^{\circ}$ east from Earth (green), 39$^{\circ}$ east of STEREO-A (red)}, and at a heliospheric distance of 0.81~AU from the Sun.  Such configuration enabled the observation of successive CMEs that caused the spread of energetic particles in the inner heliosphere. The SolO spacecraft, which was located at 115$^{\circ}$ west of Earth at a distance of 0.87 AU, was also impacted by energetic particles \citep[][not shown in the figure]{Kollhoff2021}.
Two of the observed CMEs (CME1 and CME2) directly impacted PSP \textbf{(ICME1 and ICME2)}, but only the second CME skimmed \textbf{STEREO-A. We note that during the time period of interest the position of STEREO-A does not change significantly}.  This paper focuses on the analysis of the internal structure of these two CMEs at PSP, while the STEREO-A signature will play a significant role in the characterization of ICME2 and the full reconciliation of the whole 3D scenario. 

\begin{table}[ht!]
     \centering
     \begin{tabular}{ccc}
     \hline \hline
         Event  & Start time & Location\\
             & (UT) & (deg) \\
         (1) & (2) & (3) \\
         \hline
         CME 01 & 2020-11-24 04:36 & S23E67 \\
         CME 02 & 2020-11-24 13:25 & S10E153 \\
         CME1  & 2020-11-26 21:24 & N08E102 \\
         CME2  & 2020-11-29 13:25 & S15E83 \\
     \hline
     \end{tabular}
     \smallskip
     \caption{The successive CMEs launched in the general direction of PSP during November 24--29, 2020. Columns list (1) event name, (2) date and time of first detection by the SOHO/LASCO C2 coronagraph, (3) central \textbf{CME} location.}
     \label{tab:4CMEs}
 \end{table}

\subsection{In-situ Observations}
\label{subsec:insitu}

\begin{figure}[ht!]
        \centering
            \includegraphics[width=\textwidth]{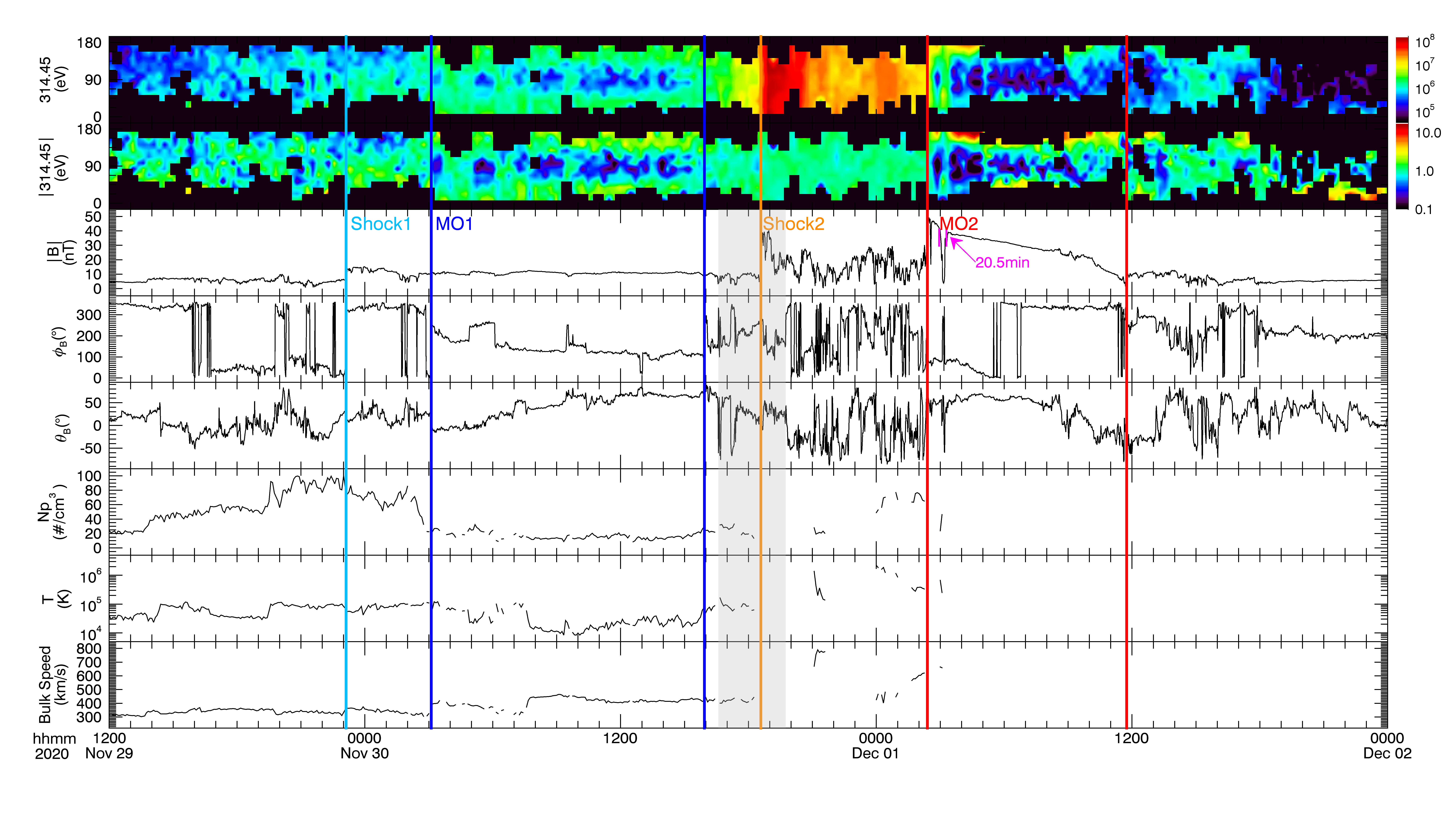}
            \caption{Overview of in-situ measurements of the November 2020 events observed by PSP during the sixth solar orbit. From top to bottom, the plots show 315 eV electron pitch angle distribution (PAD), normalized 315 eV electron PAD, magnetic field magnitude ($|B|$), magnetic field components in angular components ($\phi$, $\theta$) proton density (Np), thermal speed (Vth), and proton bulk speed. The colored vertical lines mark the boundaries of ICME1, including Shock1 (light blue) and the magnetic flux rope (dark blue), and of ICME2, including Shock2 (orange) and the magnetic flux rope (red).
            }
            \label{fig:IS}
        \end{figure}

Figure~\ref{fig:IS} displays a subset of in-situ (IS) measurements collected by the FIELDS and SWEAP instruments onboard PSP. The multipanel plot shows, from the top, pitch angle distribution of the 314 eV heat flux electrons in absolute units and normalized, the magnetic field magnitude and its spherical components, proton plasma density, thermal speed, and bulk speed. During this period, the PSP spacecraft performed several maneuvers that impeded the collection of full-cadence plasma data. Despite this, it was possible to characterize two successive ICMEs relying uniquely on the data from the FIELDS instrument.

\textbf{ICME1 was observed on November 29, 2020 at 23:07 UT}. In general, the structure is characterized by very weak magnetic field and plasma signatures. The start of the ICME1 is marked by a very weak interplanetary shock or pressure pulse (Shock1, light blue line). The magnetic observations of the Shock1 includes a very large, $\sim$270$^\circ$ magnetic field rotation (23:07:12.25--23:08:06.10) on 2020 November 29. The first half of this field rotation is magnitude preserving, while the second half (23:07:47.75--23:08:06.10) is compressive. Magnetic coplanarity calculation gives a shock normal of (0.79, 0.20, 0.58) in Radial--Tangential--Normal \citep[RTN; see e.g.][]{Hapgood1992} coordinates resulting in $\theta_\mathrm{Bn}$ = 9.1$^\circ$ as the angle between the normal and upstream magnetic field, making this a quasi parallel shock. This shock appears to be in its early stages of formation as it has no identifiable foot, overshoot, or wave activity, and the shock ramp is fairly wide at 18 seconds. In the four-hour long sheath region the magnetic field fluctuates and rotates several times with angles greater than 90$^\circ$ right before the magnetic obstacle (MO1) starts (first dark blue vertical line on November 30 at 03:21 UT). MO1 is characterized by a period of coherent change in the magnetic field magnitude for $\sim$13 hours and with a maximum of \textbf{$B_\mathrm{max}=14.0$ nT and an average of $\langle B\rangle=10.0$ nT}, twice the solar wind magnetic field strength upstream of Shock1, \textbf{$B_\mathrm{upstream}=5.5$ nT}. The magnetic field configuration exhibits a very flat and symmetric magnitude profile with a DiP of $\sim$0.5 \citep[Distortion Parameter, see][for more details]{Nieves-Chinchilla2018SolPhy}, and a smooth and small rotation in the magnetic field direction that indicates the presence of a twisted magnetic flux rope. We identified the end of MO1 (second dark blue line) a few hours before the passage of the second interplanetary (IP) shock wave (Shock2, orange line) on November 30 at 16:26 UT. Shock2 is preceded by a transition period of increase in electron flux intensity and rotation of the magnetic field vectors ($>$90$^\circ$). Despite a few rapid changes in the magnetic field direction during this period, the overall magnetic profile follows the trend marked by MO1 (see the gray area in the Figure~\ref{fig:IS}). The material measured during this period, identified as Shock2's upstream, may have originally been part of MO1 or its wake, and is being disturbed by the passage of Shock2 at the time of detection at PSP.

\textbf{ICME2 is also} preceded by an interplanetary shock (Shock2, orange line) and followed by a sheath region of magnetic field perturbations with an average magnetic field of \textbf{$\sim$11.0 nT}. This shock is much more pronounced than Shock1, with strong wave activity in the shock foot and with an initial very steep ramp (18:35:32.3--18:35:33.2) on November 30.  After this initial steep ramp, a shallower growth phase begins (till 18:35:43.05) resulting in a prolonged, 1.5-minute overshoot region. Thus, this shock also appears to be in the process of non-linear steepening. Magnetic coplanarity gives a shock normal direction of (0.33, 0.21, -0.92) in RTN coordinates and $\theta_\mathrm{Bn}$ = 24.3$^\circ$, again a quasi-parallel shock. The magnetic obstacle (MO2) starts on December 1 at 02:24 UT and ends the same day at 11:17 UT (bounded by the red vertical lines). The maximum magnetic field strength is \textbf{$B_\mathrm{max}=43.0$ nT} and the average is \textbf{$\langle B\rangle=27.0$ nT}. Note that this is almost 2.5 times the average value during the front sheath. The MO2 configuration is characterized by a declining magnetic field magnitude, with a compression at the front (DiP $\sim$0.41). The lack of plasma observations prevents us from discerning if this is a signature of distortion or expansion \citep[see][]{Nieves-Chinchilla2018SolPhy}. At the front of MO2, within the boundaries marked in Figure~\ref{fig:IS} with red lines, there is a single rotation in the magnetic field direction of more than 90$^\circ$ again, but in this case there is also a sharp intensity drop for 20 min and 29.9 seconds starting at 02:56:47 UT and ending at 03:17:17 UT on December 1 (period marked by the short pink lines in the $|B|$ panel) with a change of the magnetic field from 41.8~nT to \textbf{3.8~nT} and back to 32.3~nT. This interval is also associated with an increase in the bidirectional electrons flux intensity. The lack of plasma observations prevents us from identifying these short changes and sharp rotations in the magnetic field as magnetic-reconnection \textbf{exhaust} processes eroding the MO2 front \citep[see][]{Gosling2005, Feng2013}.

This study is focused on the analysis of the ICMEs that impacted PSP, however the full characterization \textbf{requires constraining} the forward modeling as well as the propagation models. One relevant constraint is the observations of the ICME2 by STEREO-A instruments. 
\textbf{Later in this paper, Figure~\ref{Fig:ENLIL_arrival}(b), }we will use the one-minute-averaged magnetic field observations from the IMPACT/Magnetic Field Experiment \citep[In situ Measurements of Particles And CME Transients instrument suite;][]{Acuna2008,Luhmann2008}, and the solar wind ten-minute-averaged plasma data (only available for STEREO-A) are from the Plasma and Suprathermal Ion Composition \citep[PLASTIC,][]{Galvin2008} investigation, \textbf{to constrain the CME propagation modeling by reconciling with the remote-sensing analysis.} The ICME2 at STEREO-A is identified by the Shock2 arrival at 07:30~UT on December~1, followed by an increase of the magnetic field, solar wind bulk speed, temperature, and density. However, the observations do not display any signatures of a magnetic obstacle, which may indicate that the spacecraft only crossed the flank of ICME2.

\subsection{Solar and Heliospheric Observations Between November 26--29}

To understand the origins of the ICMEs observed by PSP, we studied the source regions in extreme ultraviolet (EUV) observations and tracked the subsequent CMEs in white light (WL) coronagraph observations.  We reviewed data from the STEREO/SECCHI suite of instruments, GOES-16/SUVI, SDO/AIA, MLSO/KCor and SOHO/LASCO C2. The locations of the spacecraft during the 2020 November 26--29 time period were such that \textbf{the respective source regions were visible} on-disk in the Extreme Ultraviolet Imager (EUVI) onboard STEREO-A and at the limb (originating just behind the limb) in AIA and SUVI (see Figure \ref{fig:spacecrafts}).  Together with careful processing using advanced techniques \citep[e.g., ][]{Morgan2006,Morgan2012,Morgan2014,Alzate2021}, these observations enable a continuous view of the corona, which helps establish direct correspondence of features in EUV with those in WL.  Here, we present an overview of the source regions and corresponding activity responsible for CME1 and CME2.  However, an in-depth analysis will be presented in a subsequent study.

\begin{figure}[ht!]
        \centering
            \includegraphics[width=\textwidth]{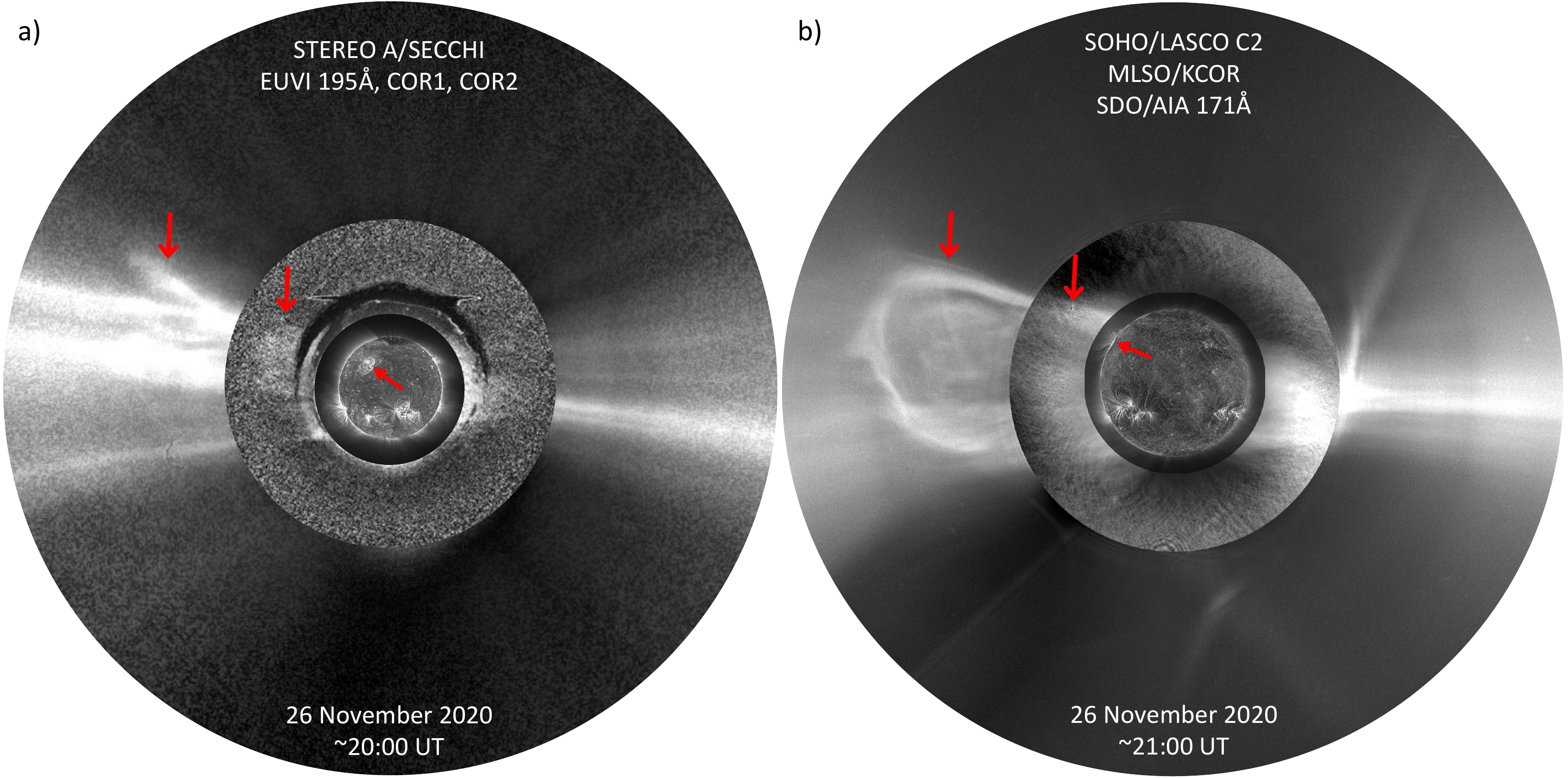}
            \includegraphics[width=\textwidth]{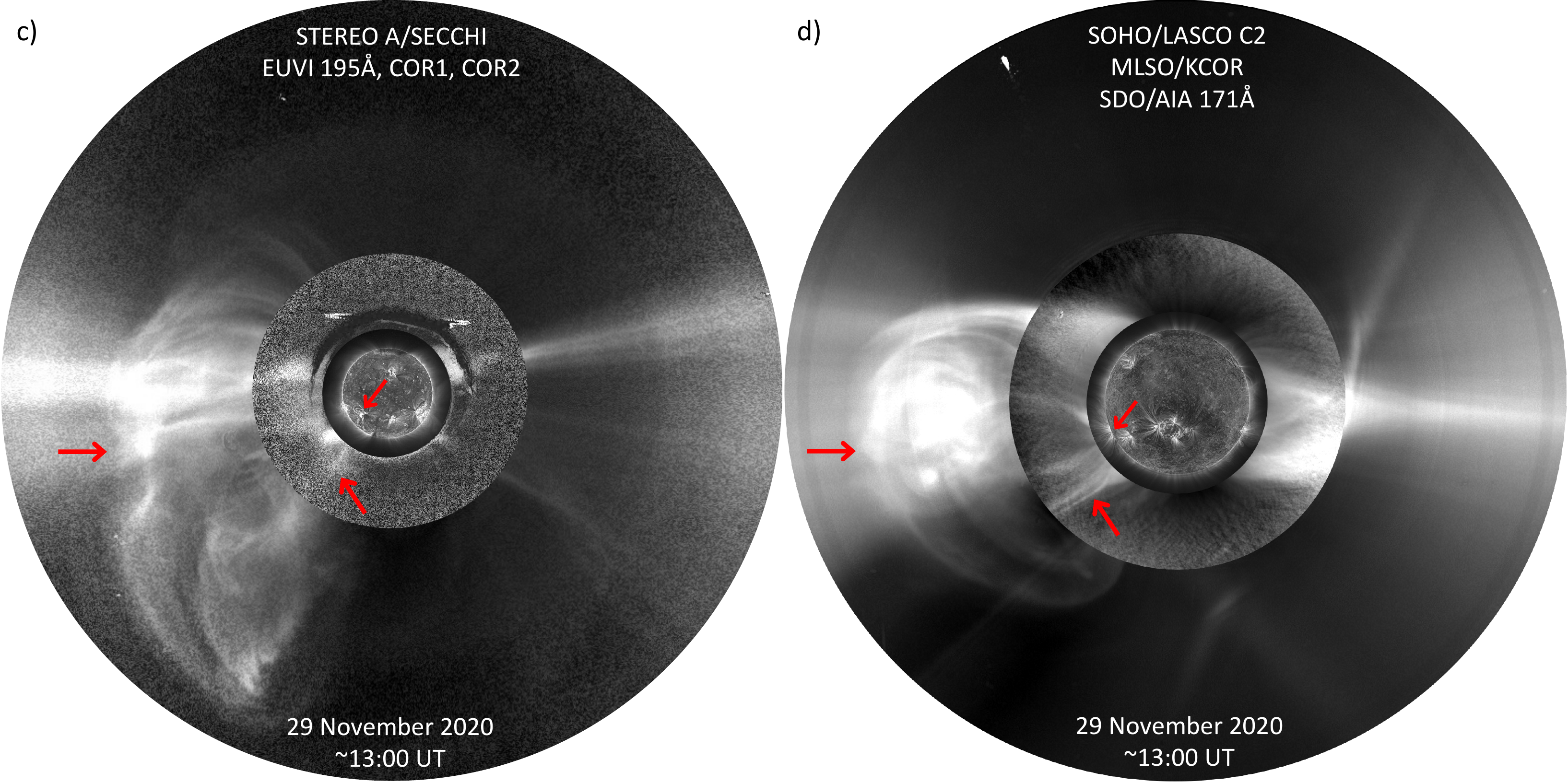}
            \caption{Source location and evolution of CME1 (a-b) and CME2 (c-d) in EUV and WL remote-sensing data.  a) November 26, 2020 STEREO-A/SECCHI composite of EUVI 195 {\AA}, COR1, and COR2 images spanning a combined FOV out to $\sim$8 $R_{S}$ as shown, with EUVI shown out to 1.4 and COR1 out to 3.0 $R_{S}$; b) November 26 composite of SDO/AIA 171 {\AA}, MLSO/KCor and SOHO/LASCO C2 images spanning a combined FOV out to $\sim$6 $R_{S}$ as shown, with AIA shown out to 1.25 and KCor out to 2.5 $R_{S}$; c) November 29, 2020 SECCHI composite spanning a combined FOV out to $\sim$10 $R_{S}$, with EUVI shown out to 1.45 and COR1 out to 3.5 $R_{S}$; d) November 29 composite of AIA, KCor and C2 images spanning a combined FOV out to $\sim$6 $R_{S}$, with AIA shown out to 1.25 and KCor out to 2.5 $R_{S}$.  Arrows indicate the source regions in the central EUV images and the CME events in the WL images.  Times shown are approximate.  Online movies of both EUV and WL images showing the evolution of each event accompany this figure.}
            \label{fig:CME_locations}
        \end{figure}

\begin{figure}[ht!]
        \centering
           \includegraphics[width=\textwidth]{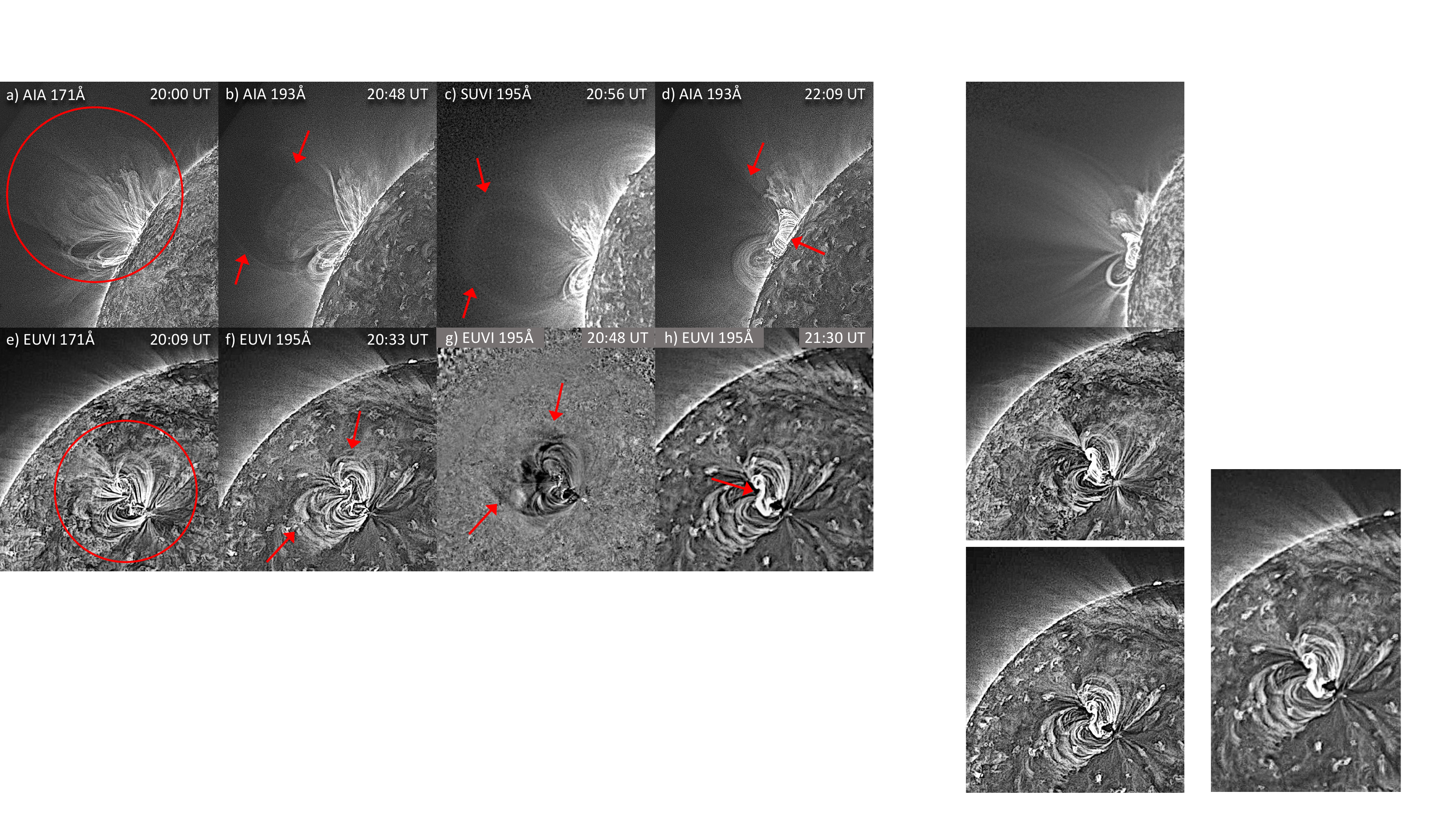}
                    
        \vspace{0.25cm}
        
           \includegraphics[width=\textwidth]{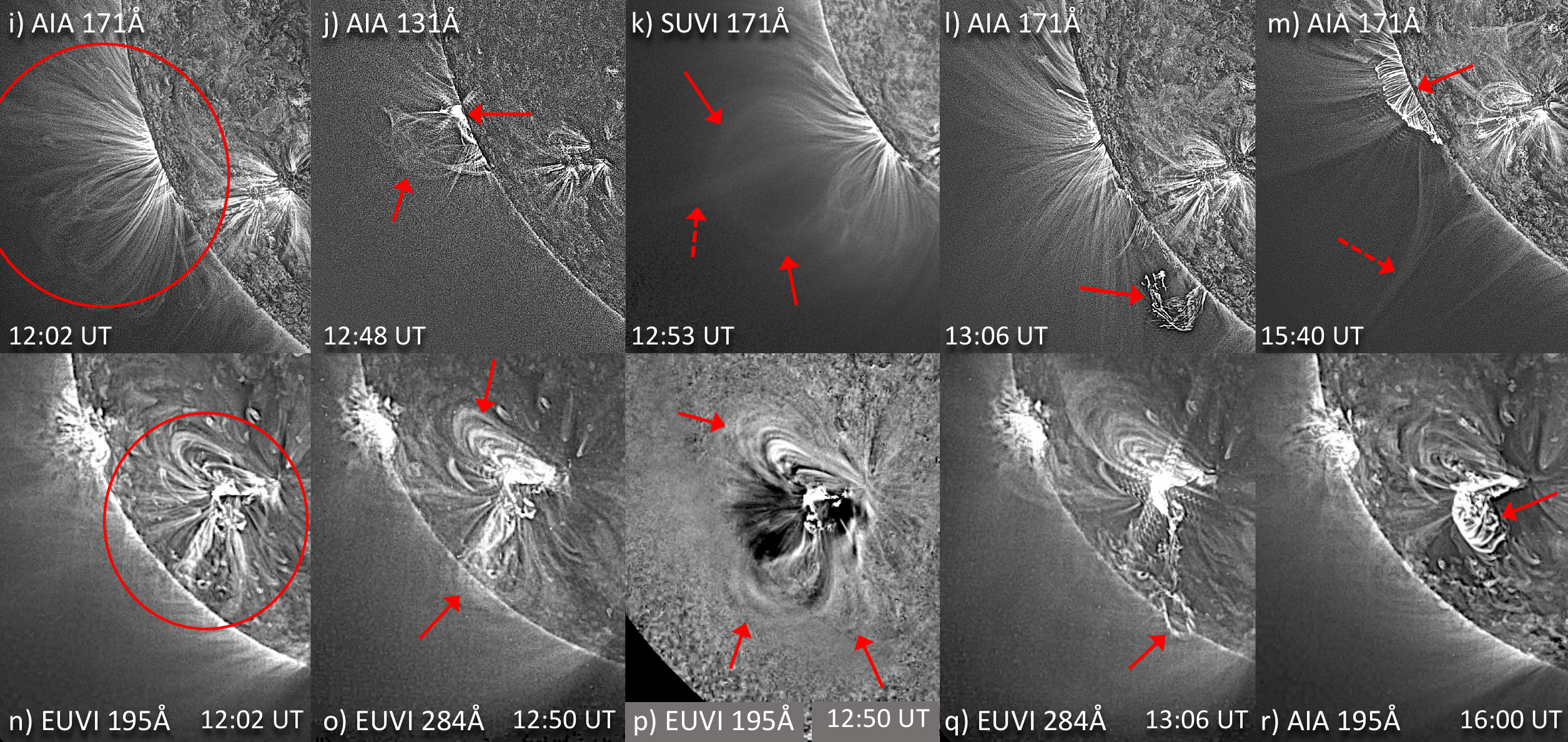}
            \caption{a--h) Time series of EUV images showing the source region and evolution of CME1 as observed along the Sun--Earth line (a--d) and by STEREO-A/SECCHI (e--h).  (i--r) Time series of EUV images showing the source region and evolution of CME2 from the Sun--Earth line (i--m) and observed by STEREO-A/SECCHI (n--r).  Online movies of the WL images showing the evolution of each event accompany this figure.   }
            \label{fig:CME-all-euv}
        \end{figure}

Figure \ref{fig:CME_locations} shows the source location and evolution of CME1 (a--b) and CME2 (c--d).  Panels a and c are composed of observations from the STEREO/SECCHI suite.  The EUVI camera observes the corona up to $\sim$1.7 $R_{S}$.  Observations with a cadence of $\sim$5 minutes in the 171, 195, and 284 {\AA} channels were used.  The SECCHI/COR1 inner coronagraph \citep{Thompson2003} observes at a $\sim$5-minute cadence and has a field of view (FOV) that extends from $\sim$1.4 out to 4.0 $R_{S}$.  The SECCHI/COR2 outer coronagraph observes the corona between $\sim$2.5 and 15 $R_{S}$ at a $\sim$60-minute cadence of polarized brightness sequences.  Panels b and d are composed of observations along the Sun--Earth line.  The LASCO/C2 instrument onboard SOHO acquires WL coronagraph observations and has a FOV spanning from 2.2 to 6.0 $R_{S}$, and a cadence of $\sim$12 minutes during the time period of interest.  The AIA instrument onboard SDO images the solar atmosphere in seven EUV channels and three UV channels out to $\sim$1.3 $R_{S}$.  For this study, images in the 131, 171, and 193 {\AA} channels were used with a reduced cadence of $\sim$24 seconds, which was enough for our purposes in this study.  The KCor instrument has a FOV from 1.05 to 3.0 $R_{S}$ and observes at a time cadence of 15 seconds.  Since KCor is a ground-based coronagraph, observations are limited to a few hours per day.  Additionally, we made use of observations from the SUVI instrument \textbf{on the GOES-16 spacecraft of the National Oceanic and Atmospheric Administration (NOAA; see Figure \ref{fig:CME-all-euv})}.  It offers a FOV of up to 1.6 $R_{S}$ on the horizontal and 2.3 $R_{S}$ on the diagonal and observes at a regular cadence of four minutes.  We used images in the 171 and 195 {\AA} channels.

On 2020 November 26, CME1 erupted from the northeast region (see Table~\ref{tab:4CMEs} for coordinates) indicated in Figure \ref{fig:CME_locations}a and 3b and was first seen in the LASCO C2 FOV at 21:12 UT.  Prior to entering the FOV, CME1 was preceded by a series of very faint outflows as the streamer slowly began to expand around 20:36 UT.  CME1 overtook the FOV by 22:36 UT and was followed by post-CME outflows.  At 1:25 UT, a small and faint structure appeared tethered to CME1 and was followed by more outflows until the end of the 2020 November 27.  CME1 itself exhibits a bright leading front, followed by a cavity and a bright central core \citep[thus featuring a classic ``three-part'' morphology;][]{Illing1985,Chen2011,Vourlidas2013}. No signs of a preceding shock can be discerned in the WL images of CME1. The apparent radial velocity of CME1 in LASCO C2 is $\sim$524 $kms^{-1}$.  In ancillary observations by the ground-based coronagraph KCor, the streamer in the northeast region undergoes subtle reshaping and expansion between 22:02 and 22:36 UT on 2020 November 26.  Entrained within the streamer is a ``triangle-like" shape protruding from the inner FOV up to $\sim$1.5 $R_{S}$.  This structure matches the shape and location of the erupting loop system in EUV that formed the core of CME1 before it propagated outward in the extended LASCO C2 images.  

Figure~\ref{fig:CME-all-euv} shows observations by AIA and SUVI of the \textbf{CME1's} source region at the limb at approximately 30$^{\circ}$ counterclockwise from the north (Panels a-d in Figure \ref{fig:CME-all-euv}). It exhibits a complex system of loops that develop \textbf{into the shell of} CME1 (panel a).  In AIA, loops begin to expand at 20:25 UT and continue to expand until an eruption begins at around 20:48 UT (panel b) and leaves the AIA FOV (1.3 $R_{S}$) at 21:25 UT.  Also at this time, the remaining loops appear to twist and form the ``triangle-like" shape observed in KCor.  A post-eruption set of bright loops begins to form at 21:30 UT at the onset of a flare, followed by flare loops (panel d).  The extended FOV of SUVI provides a better view of the loop system, outflows and the surrounding structure.  A series of faint pre-eruption outflows begin at 13:00 UT and continue after the main eruption until the bright post-flare loops are formed.  

CME1 first enters the COR1 FOV at 21:25 UT and can be tracked for a few frames until 22:15 UT.  Because of instrument noise and deterioration, it is difficult to characterize small-scale structures within CME1.  Only streamer brightness enhancements are obvious.  In COR2, CME1 is first visible at 22:08 UT.  CME1 exits the COR2 FOV beginning at 02:08 UT on the 2020 November 27.  The source region in EUVI is on the disk near the East limb (panels e-h in Figure \ref{fig:CME-all-euv}, see Table~\ref{tab:4CMEs} for coordinates).  Because of the on-disk location, we can study the evolution of the region.  In the 195 {\AA} channel, the system of loops open and erupt beginning at 20:33 UT (panel f). The pre- and post-eruption loops are visible, but what is most distinguishable is the bright ``S-shape" loop structure, or sigmoid \citep[e.g.][]{Green2007}, best seen at 21:30 UT (panel h).  In a time series (not shown) of bandpass-filtered images \citep[][]{Alzate2021}, the loops open and the eruption begins at 20:48 UT (panel g), with the sigmoid best seen at 21:48 UT.  The associated wave is distinctly seen on the solar disk beginning at 20:55 UT and overtakes a large region at approximately 21:30 UT.  The wave appears to trigger the first of many outflows originating from the active region that can be seen at the limb below the equator (115 degrees counterclockwise from the north).  This is the source region of CME2 described below.

On 2020 November 29, CME2 erupted from the southeast region indicated in Figure \ref{fig:CME_locations}c and d.  The CME is first visible in the LASCO C2 FOV at 13:25 UT, though there is a preceding data gap of 36 minutes, equivalent to three LASCO C2 frames. The presence of a shock in the WL images of LASCO C2, COR1, and COR2 is straightforward. Observations from KCor on 29 November began at 17:49 UT, after CME2 had left the LASCO C2 FOV.  However, because this CME appears large in LASCO C2, its following imprint is visible in KCor (Figure \ref{fig:CME_locations}d).  A similar imprint is left behind after CME2 exits the FOV of COR1 and COR2.  Similar to CME1, the source region from which CME2 originated (panels i--m in Figure \ref{fig:CME-all-euv}) is characterized by a complex system of loops that begin expanding at 12:40 UT.  The flare is best seen in the 131 {\AA} channel (panel j). The eruption then follows at 12:50 UT and plasma material outflows southward from the active region (panel l).  After the main eruption, material continues \textbf{to be expelled} towards the south and bright post-flare loops form (panel m).  In the extended FOV of SUVI, the full extent of the loop system is visible out to $\sim$1.9 $R_{S}$ (panel k).  The outer set of loops begin to open at 12:53 UT.  Inside the loops is a radial structure that protrudes from the source region and cuts through the center of the loops as indicated in panel k. 

In COR1, restructuring of the streamer occurs over a period of a few hours leading up to CME2, which is first seen at 13:00 UT.  CME2 overtakes the FOV by 13:30 UT before the bubble is fully evacuated by 13:50 UT. In COR2, CME2 is first seen at 13:24 UT.  It expands quickly with an apparent radial velocity of  $\sim$1275 km s$^{-1}$ and overtakes the FOV by 14:24 UT.  Post-CME outflows are visible until the end of the 29$^{th}$.  In EUVI 195 {\AA} images, the region of interest is highly active (panels n--r in Figure \ref{fig:CME-all-euv}).  The wave associated with the main eruption of 26 November in the northern region (CME1) reached the source region of CME2 and triggered a series of outflows up until the flare event that triggered the CME.  Pre-eruption events begin with a large outflow seen at 08:10 UT and directed southward.  The main eruption begins slowly at 12:50 UT (panel o), but quickly picks up speed and erupts at 13:05 UT with a post-eruption, loop-like twisting outflow (panel q).  Bright post-eruption loops begin to form (panel r).  In bandpass-filtered images \citep[][]{Alzate2021}, outflows and inflows are visible starting at 08:20 UT and appear to trigger the large main eruption event, which begins at 12:25 UT.  Panel p shows an example of a later frame where the outer post-eruption loops are indicated with arrows.  The associated wave expands and overtakes most of the east around 13:00 UT while triggering small eruptions nearby.

\subsection{Multispacecraft Radio Observations}

CMEs and solar flares are associated with intense radio signals in a wide range of frequencies, in particular with type II and III bursts. They are generated via the plasma emission mechanism, when electron beams interact with the ambient plasma triggering radiation at the plasma frequency $f_{\rm{p}}$ (the fundamental emission), or at its second harmonic $2f_{\rm{p}}$ (the harmonic emission). As the electron beams propagate outward from the Sun, radio emissions are generated at progressively lower frequencies corresponding to a decreasing local plasma density \citep{1950AuSRA...3..541W,1958SvA.....2..653G,1998ApJ...503..435E}. Type II bursts are triggered by electron beams accelerated at the shock fronts driven by fast moving CMEs, while type III bursts are a consequence of impulsively accelerated electrons associated with solar flares \citep{1998GeoRL..25.2493R,2000GeoRL..27..145G,2016ApJ...823L...5K}. Type II bursts are intermittent with short periods of radio enhancements, which are usually related to CME--CME and/or CME--streamer interactions \citep[e.g.][]{2014ApJ...791..115M}.

\begin{figure}
\centering
\includegraphics[width=0.45\textwidth]{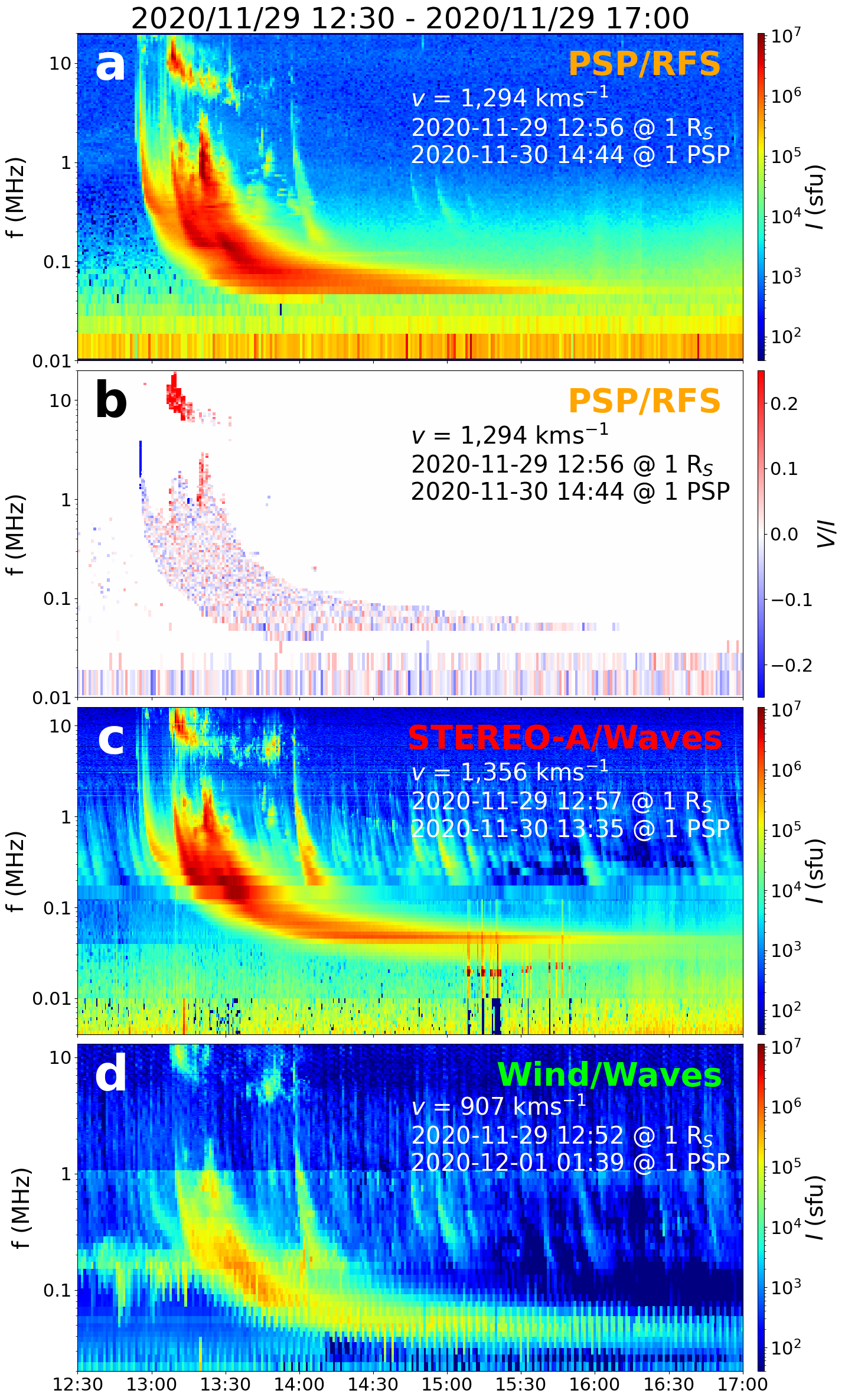}
\caption{Radio measurements of type II and type III bursts on November 29 from 12:30 UT to 17:00 UT. (a) Radio flux density $I$ at PSP ($1$~sfu~=~$10^{-22}$W/m$^2$/Hz). (b) Relative circular polarization $V$/$I$ (positive values indicate LHC polarization). (c,d) Radio flux density $I$ at STEREO-A and Wind. Results of type II burst linear fit are shown in all panels. The intensity threshold ($I>10^5$~sfu) has been applied on the (b) to suppress the background. Included in each panel is the time of the Type II observation and the estimated arrival time to PSP. }
\label{fig:radio}
\end{figure}

CME2 was accompanied by type II and III bursts detected by PSP/RFS, STEREO-A/Waves, and Wind/Waves \citep{Wilson2021,2016SSRv..204...49B, 2008SSRv..136..487B, 1995SSRv...71..231B} in the period of November 29 12:30- 17:00 UT (Figure~\ref{fig:radio}). A complex type III occurred around 12:55~UT, and it lasted until $\sim$16:00~UT. Langmuir waves were observed by STEREO-A between 15:10~UT and 15:50~UT, indicating that a flare-associated electron beam intersects the STEREO-A spacecraft (Figure~\ref{fig:radio}c). The complex type III was followed by an intermittent type II burst between 13:05~UT and 13:50~UT in a frequency range from 20~MHz down to 4~MHz (Figure~\ref{fig:radio}).  It corresponds to radial distances of the CME2-driven shock wave between $\sim 1.8\,R_{S}$ and $\sim 3.7\,R_{S}$ assuming the fundamental emission and the electron density model of \citet{Sittler1999}. The type II burst could have been generated either by interaction of the CME2-driven shock wave with the base of CME1 legs, or a nearby streamer. This event is the first well observed type II burst by three spacecraft located at different longitudes and latitudes. 

We analyzed the frequency drift of the type II burst around 13:10~UT to estimate kinematics of the CME2-driven shock wave in the corona. Specifically, we identified peak fluxes ranging between $7$~MHz and $13$~MHz for each spacecraft separately. Next, we converted frequencies of the type II burst $f$ to radial distances $r$ using the density model of \citet{Sittler1999}. Using a linear fit we derived the speed of the CME2-driven shock to be $1,294$~kms$^{-1}$, $1,356$~kms$^{-1}$, and $907$~kms$^{-1}$ for PSP, STEREO-A, and Wind, respectively. While frequency drift results from PSP and STEREO-A observations are comparable, the obtained speed for Wind is significantly different. It can be explained by different radio propagation effects as the signal suffers by refraction in density gradients and scattering by density inhomogeneities. \citet{2019ApJ...882...92K} showed that type II bursts are more likely to have a source region situated closer to CME flanks than CME leading edge regions. In the case of PSP and STEREO-A, CME2 propagates roughly between the two spacecraft, while Wind is located nearly perpendicular. As the type II burst is more likely to be generated near CME2 flanks, Wind observed the radio emission with smaller propagation effects, when compared to the two other spacecraft.

Assuming a constant speed for the CME2-driven shock, we calculated the onset ($r=1$~$R_{S}$) and PSP arrival ($r=174$~$R_{S}$) times (Figure~\ref{fig:radio}). The obtained onset time for Wind (2020 November 29 12:52~UT) is consistent with observed ground-based radio measurements. The same type II burst was detected at $80$~MHz ($\sim 1.17$~$R_{S}$) around 2020 November 29 12:54~UT
\citep[Figure~4 of][]{Kollhoff2021}. Finally, thePSP arrival time based on Wind frequency drift (2020 December 01 at 01:39~UT) is close to the actual time of the CME1--CME2 interaction analyzed in this study (2020 December 01 at 02:20~UT, see Figure~\ref{fig:DBM} below).

Moreover, the type II burst exhibited significant left-hand circular (LHC) polarization \textbf{which has never been detected} before in this frequency range (Figure~\ref{fig:radio}b). The PSP/RFS is the first space-based radio receiver that allows us to investigate polarization properties of incident radio waves up to $20$~MHz. STEREO/Waves and Wind/Waves provide direction-finding measurements only up to $2$ and $1$~MHz, respectively. \citet{2020ApJS..246...49P} analyzed several type III bursts with significant LHC polarization observed by PSP/RFS in 2019. This strong polarization is surprising, as solar radio emissions at frequencies below ~20 MHz rarely show signatures of circular polarization \citep{Cecconi2019}. When it is present, strong circular polarization is often interpreted as evidence that the emission occurs at the fundamental of the plasma frequency, and not the harmonic. In the presence of a magnetic field, electromagnetic waves can be generated in O-mode and X-mode, which have opposite senses of circular polarization. At the fundamental, only O-mode emission can escape the source region and reach the observer, while at the harmonic, both O-mode and X-mode waves can escape. This implies that fundamental emission should be more strongly polarized than harmonic. In the case of Type IIIs, this is well established with observational results \citep{Dulk1980}.

\textbf{The handedness of radio burst emission can also be used to probe the magnetic field conditions in the source region, and, in this specific case, help determine the radio emission source location. Later in this paper, the Figure 10(c) will illustrate the geometry of the ICME interaction, which will be described in detail in Sections 3 and 4. This shows the two legs of ICME1, a ``north'' leg extending towards the positive $z$ and negative $x$ directions, and a ``south'' leg, which is close to the $x$ axis. From the information in Table 3, we can determine that the polarity of the central magnetic field is positive (outward from the Sun) in the south leg of the ICME, and negative (towards the Sun) in the north leg.}

\textbf{In this geometry, PSP measurements of radio emission from the south leg correspond to radio waves which propagate parallel to the central field, while waves from the north leg propagate antiparallel to the field. This alignment determines the sense of circular polarization \citep{Thejappa2003}, with LHC polarization in the parallel (south leg) case, and RHC polarization in the antiparallel (north leg) case. The LHC-dominated emission in Figure~\ref{fig:radio} therefore indicates that the observed type II emission is predominantly generated in the south leg of ICME1, as ICME2 passes through and interacts with ICME1.}

\section{3D Reconstruction of the CMEs/ICMEs} \label{sec:Recons}

In this section we focus on the reconciliation in the 3D reconstructions of the CME/ICME among the three perspectives of IS, remote-sensing observations, and ENLIL numerical modeling. This exercise is challenged by the lack of plasma in-situ observations and the limitation on the forward modeling techniques to fully characterize the remote sensing observations and the propagation of the ICME in space. Thus, the 3D reconstruction of the CMEs in the heliosphere is set out by the synchronization of the remote-sensing reconstruction with the PSP and STEREO-A in-situ ICME observations. In this section we will first explore the 3D reconstruction based on remote-sensing observations to later reconcile with the propagation and in-situ modeling.

\subsection{3D Forward Modeling Reconstruction of the CMEs}
\label{sec:RS_Recons}

To determine three-dimensional kinematic and morphological properties of the CMEs under study, we make use of the Graduated Cylindrical Shell \citep[GCS;][]{Thernisien2009} forward model. The model relies on the manual fitting of an ad-hoc 3D geometrical figure (the GCS) to the outline of the projected white-light CME, as simultaneously viewed from coronagraphs from different vantage points. For the case of the CMEs under analysis, we used images from STEREO-A SECCHI/COR1 and COR2 together with those from SOHO/LASCO C2 and C3. These pairs of nearly simultaneous images were fit to the GCS figure at various points in time, so as to characterize the evolution of their key 3D parameters. The fitting involves careful iterative adjustment of the GCS shape, by setting different values to the six parameters that define the 3D CME morphology. This is performed until the best visual match is simultaneously achieved in every pair of images exhibiting the projected CME. The outcome of the process is a set of parameters that describe the dimensions and location of the CME at a specific time: latitude $\theta$, longitude $\phi$, tilt $\gamma$, height $h$, aspect ratio $\kappa$, and half angle \textbf{$\alpha$} \citep[see][for further description of the model parameters]{Thernisien2006}.

\begin{figure}[ht!]
\centering
\setlength{\unitlength}{1cm}
    \begin{picture}(17.8,13.35)
        \put(0,8.9){\includegraphics[width=4.45cm]{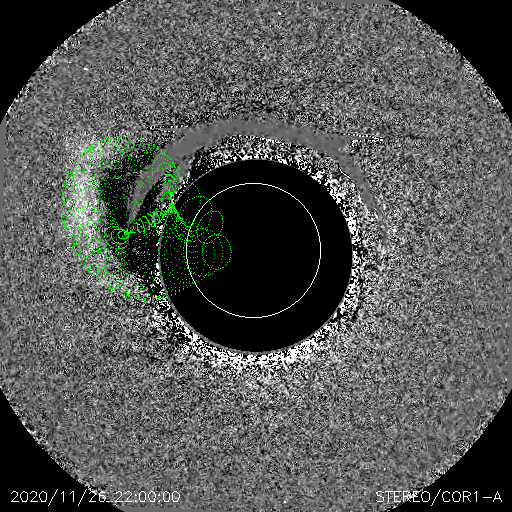}}
        \put(4.45,8.9){\includegraphics[width=4.45cm]{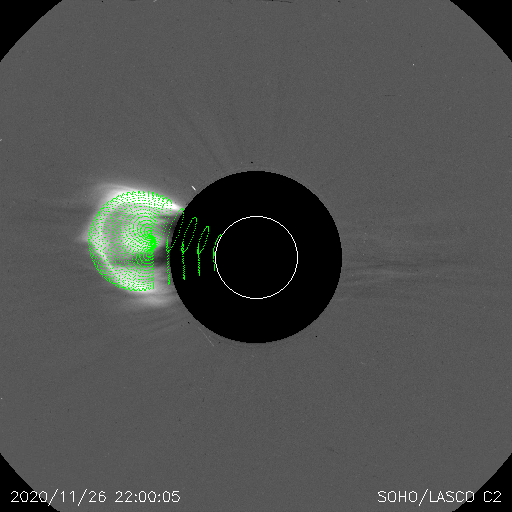}}
        \put(0.15,12.8){\textcolor{white}{\Large \sffamily a)}}
        \put(0,4.45){\includegraphics[width=4.45cm]{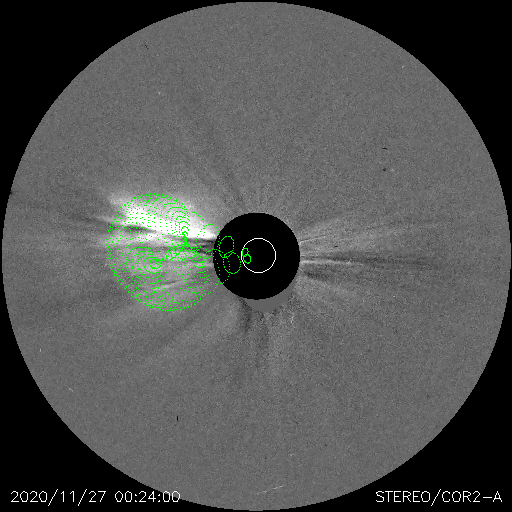}}
        \put(4.45,4.45){\includegraphics[width=4.45cm]{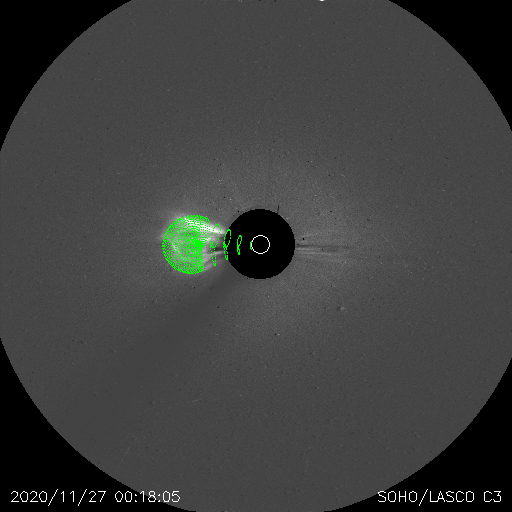}}
        \put(0,0){\includegraphics[width=4.45cm]{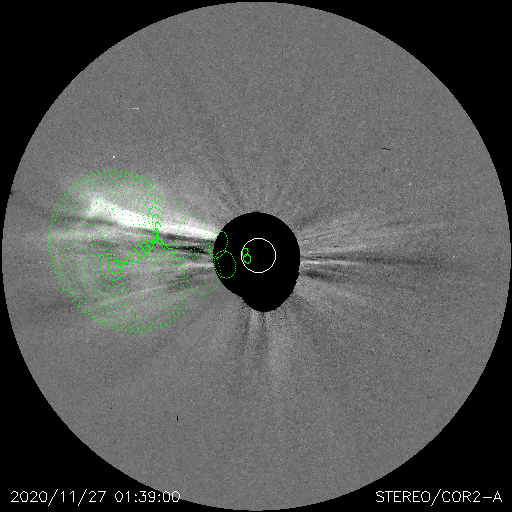}}
        \put(4.45,0){\includegraphics[width=4.45cm]{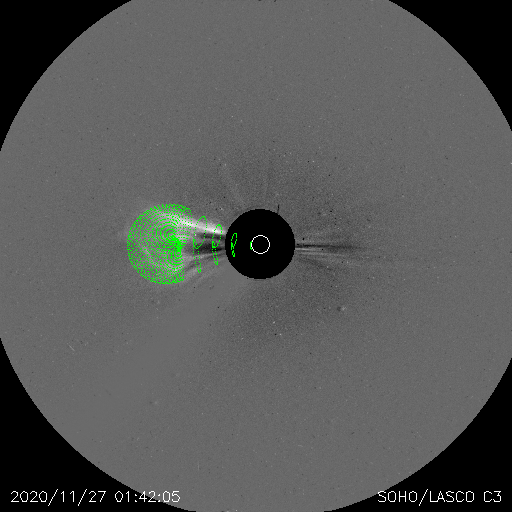}} 
        \put(9,8.9){\includegraphics[width=4.45cm]{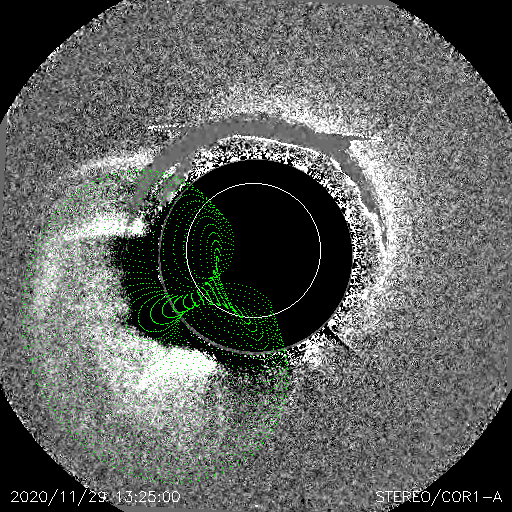}}
        \put(9.15,12.8){\textcolor{white}{\Large \sffamily b)}}
        \put(13.45,8.9){\includegraphics[width=4.45cm]{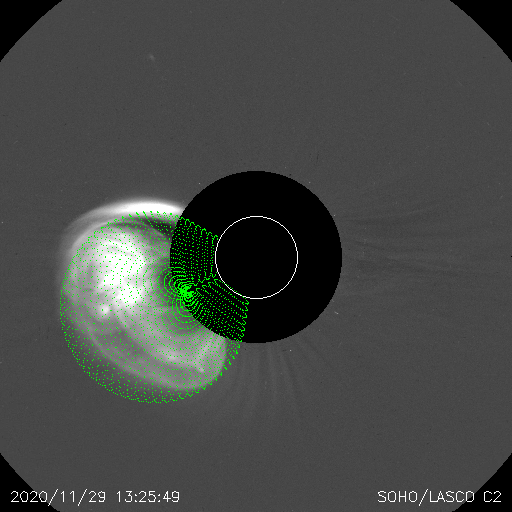}}
        \put(9,4.45){\includegraphics[width=4.45cm]{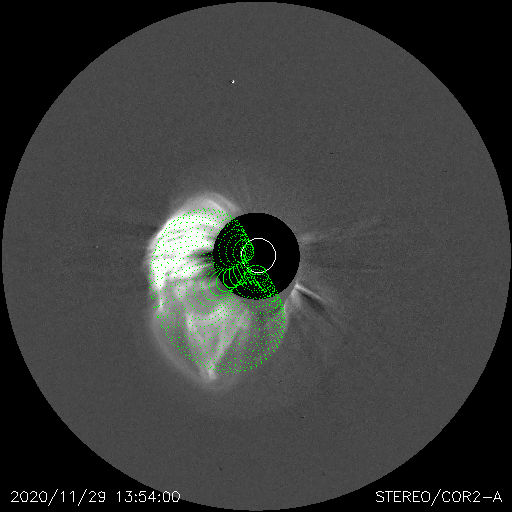}}
        \put(13.45,4.45){\includegraphics[width=4.45cm]{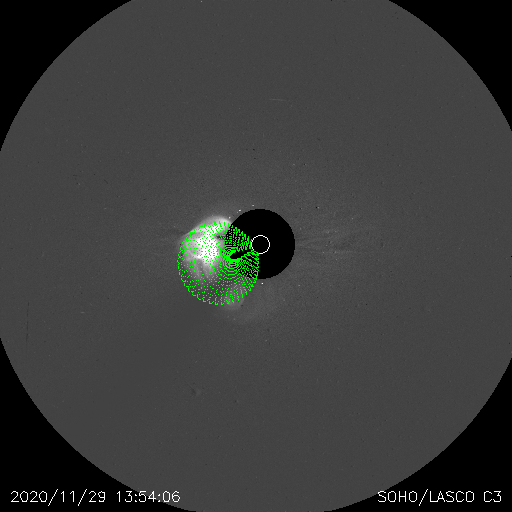}}
        \put(9,0){\includegraphics[width=4.45cm]{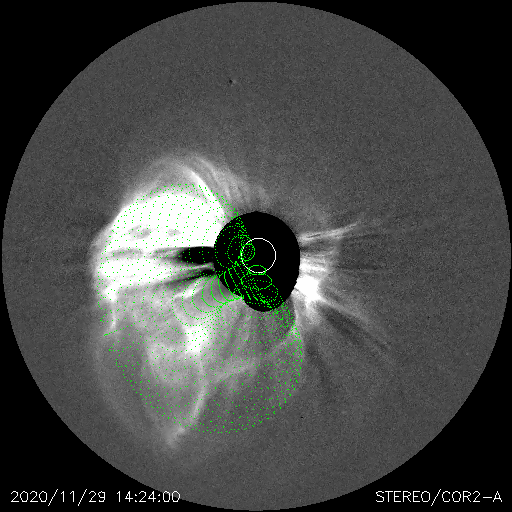}}
        \put(13.45,0){\includegraphics[width=4.45cm]{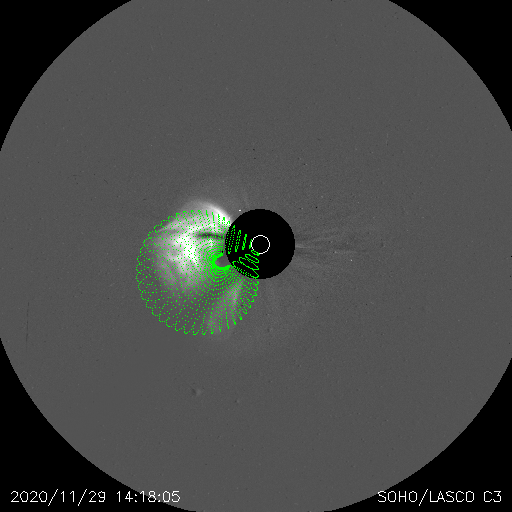}}
    \end{picture}
    \caption{GCS model reconstructions (green meshes) superimposed on a) CME1, b) CME2. The \emph{left column} in a) and b) is the view from coronagraphs onboard STEREO-A and the \emph{right column} is the view from SOHO/LASCO. Time increases from top to bottom. A pre-event image has been subtracted to all frames to increase the contrast of the CMEs with respect to the background. }
    \label{fig:GCS_CMEs}
\end{figure}

The GCS fits at three different times are shown as green meshes superimposed on white-light images of CME1 (a) and CME2 (b) in Figure \ref{fig:GCS_CMEs}. Time increases from top to bottom. The left column in (a) and (b) shows the view from STEREO-A, and the right column in (a) and (b) shows the view from SOHO. The final parameters that define the 3D shape of the CMEs, used as input for the ENLIL model runs, are summarized in Table \ref{tab:GCSENLIL}. Column (1) lists the CME number, columns (2) through (6) list the GCS parameters, and columns (7) through (11) list other parameters arising from the GCS parameters, which are required as inputs for the ENLIL model runs (see Section \ref{Sec:enlil}). Note that Table \ref{tab:GCSENLIL}, besides listing parameters corresponding to CME1 and CME2, also displays those parameters deduced for CME 01 and CME 02. These occurred prior to CME1 and CME2, thus they are important for setting the prevailing solar wind conditions during the propagation of the CMEs in this study (see next subsection). The leading edge (LE) radial propagation speed (column 10) of the CMEs has been estimated from the GCS fits at various times, while the time when the LE is at 21.5 $R_{S}$ (column 11) is deduced from it. Note that, although Figure \ref{fig:GCS_CMEs} displays GCS fits at three times, we have performed the fitting process for all available nearly-simultaneous image pairs showing the CMEs. 

When comparing parameters obtained from the first analyzed time with those a couple of hours later, a slight equatorward and eastward deflection was evident for CME1. No changes in tilt, aspect ratio, or half angle were noticeable. The resulting height--time profile for CME1 was approximately linear, yielding a speed of $\sim$580 km s$^{-1}$.

For CME2, there was a slight equatorward and westward deflection, with no apparent changes in tilt or aspect ratio, though the half angle had to be increased to better match the expanding CME. In fact, CME2 exhibited a behavior that departed from a typical CME, with its southern portion expanding to a greater extent and a leading edge flatter than that modeled by the GCS shape. The latter is particularly noticeable in SOHO/LASCO images within which the CME propagates close to the sky plane.  Three-dimensional reconstruction of CME2, on the basis of the available white-light views, was non-trivial. A process of reconciliation between GCS fits, ENLIL simulations, and in-situ reconstructions had to be pursued to minimize uncertainties in the 3D parameters. Three different GCS fits were tried, with varying results when used as input for the ENLIL runs. The first and second options led to nearly linear height--time profiles, with speeds surpassing 1900 km s$^{-1}$, yielding wrong arrival times in ENLIL with respect to the in-situ observations. The third option led to the best agreement between ENLIL simulations and in-situ reconstructions as well as observations, and is therefore the one used for CME2 and shown in the last row of Table \ref{tab:GCSENLIL}. According to the third set of GCS parameters, the CME2 leading edge (LE) initially exhibits a speed of $\sim$2000 km s$^{-1}$, but quickly decelerates to $\sim$1780 km s$^{-1}$ by the time it reaches the outer edge of the FOV. The three GCS options slightly differ in the values of all parameters: the first option is similar to the third one, but with a larger $\kappa$ (0.47 vs. 0.45) and a smaller $\alpha$ ($32^{\circ}$ vs. $37^{\circ}$). The second option involved a smaller $\kappa$ of 0.40, a tilt $\gamma=-80^{\circ}$ to be in accordance with the inclination of the post-eruptive loops, and a more eastward $\phi$ of $270^{\circ}$. Under the conditions of the second option, the value of $\phi$ implied that CME2 did not reach STEREO-A. This exercise of trying different GCS fit parameters is another example of how relatively small variations in the parameters may have a great impact on the simulations at large distances \citep[e.g.][]{Mays2015}. Note the upper and lower limits of the GCS parameters considered by these three options: $-19^{\circ}<\theta<-15^{\circ}$, $270^{\circ}<\phi<277^{\circ}$, $-70^{\circ}<\gamma<80^{\circ}$ (i.e. $30^{\circ}$ difference), $0.40<\kappa<0.47$, and $32^{\circ}<\alpha<37^{\circ}$.

\subsection{CME Propagation and Reconciliation with \textbf{Remote-sensing and In-Situ Observations}}
\label{Sec:enlil}

\begin{deluxetable}{ccccccccccc}
\tablecaption{CME parameters from GCS for ENLIL model\label{tab:GCSENLIL}}
\tablehead{
\colhead{CME} &  \colhead{Lat $\theta$} & \colhead{Lon $\phi$} & \colhead{Tilt $\gamma$}& \colhead{Ratio $\kappa$}& \colhead{Half angle $\alpha$}& \colhead{Rmaj}& \colhead{Rmin}& \colhead{Tilt2} & \colhead{LE Speed} & \colhead{Time @ 21.5 R\textsubscript{$S$}}\\
\colhead{event} & \colhead{(deg)} & \colhead{(deg)} &\colhead{(deg)} & \colhead{(-)} & \colhead{(deg)}& \colhead{(deg)}& \colhead{(deg)}&\colhead{(deg)} & \colhead{(km/s)}&\colhead{(UT in 2020)}
}
\decimalcolnumbers
\startdata
01 & -23 & -67  & -21 &0.20 &15  & 27 & 12 & -21 & 820 & 24-11 \hspace{1mm} 10:22 \\
02 & -10 & -153 &-12  &0.40 &30  & 54 & 24 & -12 & 1141 & 24-11 \hspace{1mm} 16:20 \\
\hline
1  & 8   & -102 &-60  &0.25 &12  & 26 & 14 & -60 & 576 & 27-11 \hspace{1mm} 03:56 \\
2  & -15 & -83  &-70  &0.45 & 37 & 64 & 27 & -70 & 1780 & 29-11 \hspace{1mm} 15:15 \\
\enddata
\tablecomments{Column 1: CME number, CME 01 and CME 02 erupted prior to CME1 and CME2. Columns 2 through 6: GCS output parameters at the last analyzed time. Column 7: Face-on CME half-width, calculated by adding the half-angle (6) to the edge-on CME half-width (8). 
Column 8: Edge-on CME half-width, calculated using arcsin(ratio) (5).
Column 9: Angle of the CME with the solar equator (CCW positive).
Column 10: LE CME speed at 21.5 $R_{S}$. 
Column 11: Time of CME arrival at 21.5 $R_{S}$ (ENLIL inner simulation boundary).}
\end{deluxetable}

\begin{figure}[ht!]
\centering
        \includegraphics[width=0.9\linewidth]{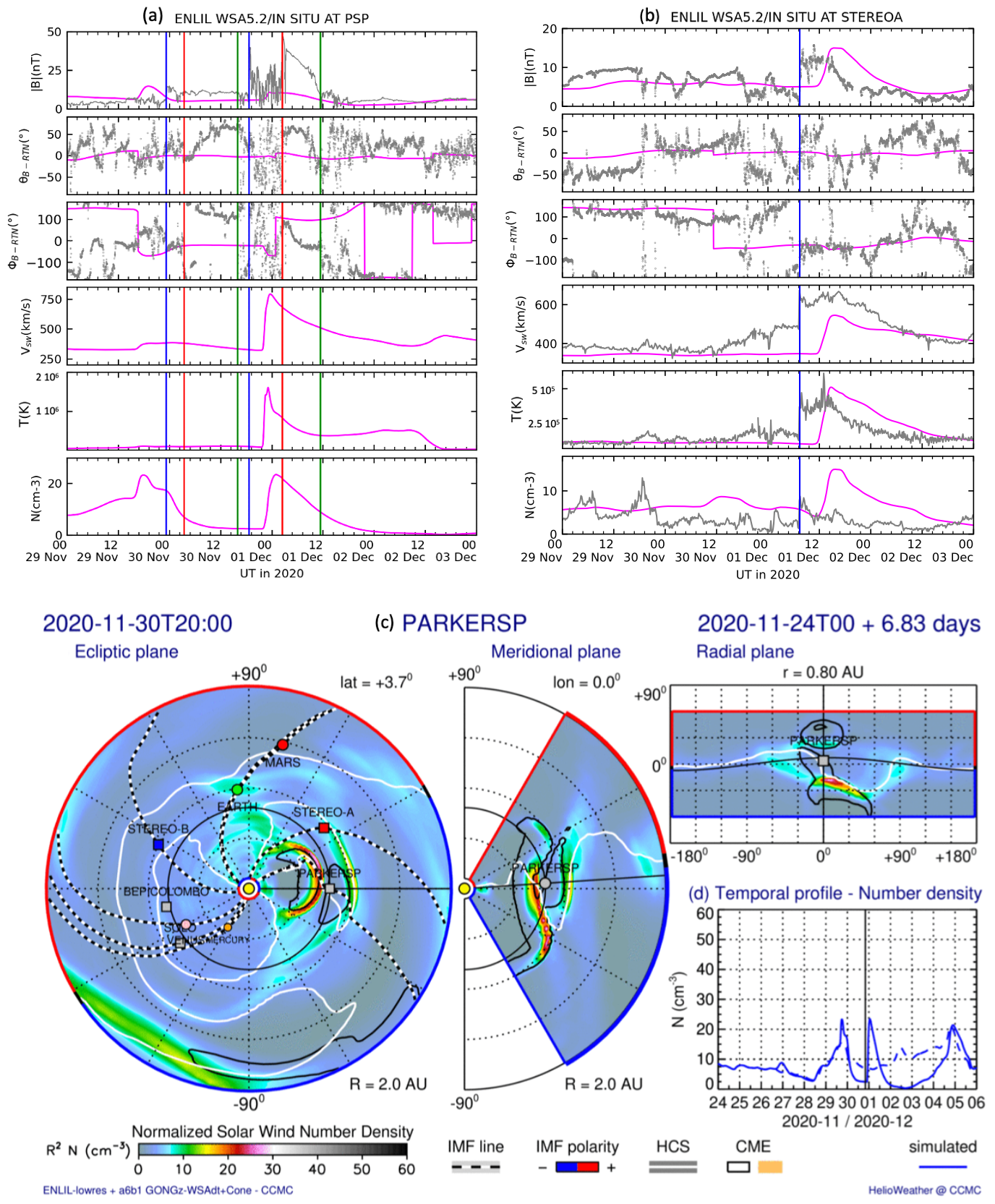}\\
        \caption{ENLIL simulation results. Top: In-situ observations at the locations of PSP (a) and STEREO-A (b), over-plotted with the ENLIL results (pink line). Blue vertical lines indicate the IP shock arrival and red and green lines respectively mark the start and end of the MO.  Bottom: Snapshot of the density contour plot from the ENLIL simulation at the arrival of ICME2 driven-shock at the location of PSP. It shows three different views, the ecliptic plane (left), meridional plane (center), and radial plane (right). The black contours track the ICMEs, as they are set as regions where the density ratio to the background is higher than a certain threshold. The black and white dashed lines represent the IMF lines and the white lines show the heliospheric current sheet, which divides regions with opposite magnetic polarity, shown in blue (negative) or red (positive) on the outer edge of the simulation area. }
        \label{Fig:ENLIL_arrival}
    \end{figure}

    \begin{figure}[ht!]
    \centering
        \includegraphics[width=0.925\textwidth]{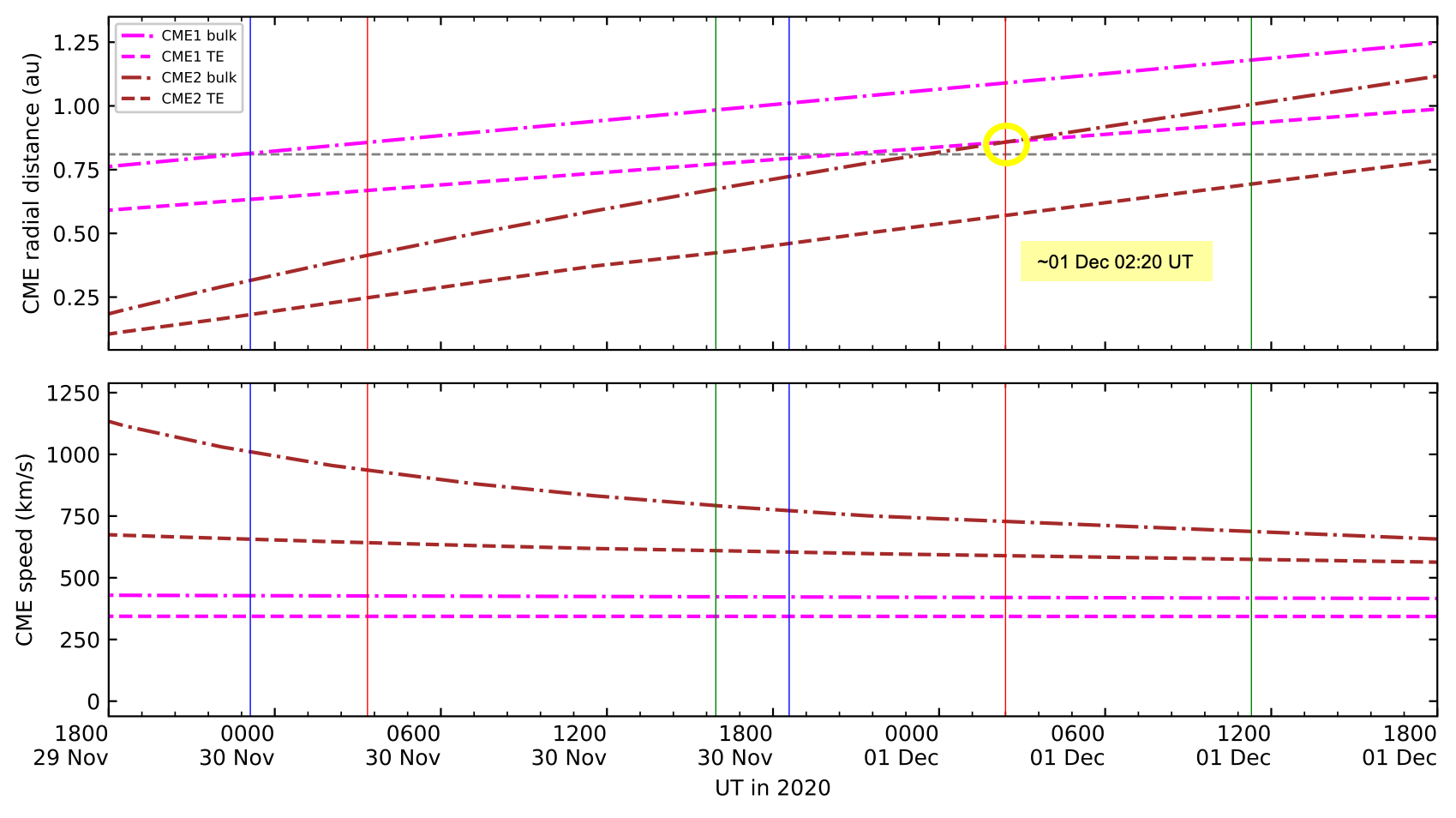}\\
        \caption{CME radial distance (top) and speed (bottom) given by the DBM model. The magenta and brown lines correspond to CME1 and CME2, respectively. The dash-dotted and dashed lines represent the bulk core and the TE of the CME, respectively. The vertical lines indicate the shock (blue), and the start (red) and end (green) of the MO observed in-situ at the location of PSP, located at 0.81 au, indicated by the horizontal grey dashed line. The yellow circle indicates the start time of the interaction between ICME1 and ICME2 as derived from the model.}
        \label{fig:DBM}
    \end{figure}

In this section, we modeled the propagation of the CMEs during the period of November 24 to December 6 to estimate the evolution of CME1 and CME2 at the location of PSP using two complementary models, the ENLIL model and the Drag-Based model \citep[DBM,][]{Vrsnak2013}.

The WSA-ENLIL+Cone model \citep{Odstrcil2004} is a global 3D magnetohydrodynamic (MHD) model\footnote{\url{https://ccmc.gsfc.nasa.gov/models/modelinfo.php?model=ENLIL\%20with\%20Cone\%20Model}} that characterizes the heliosphere outside 21.5 $R_S$. ENLIL uses a time-dependent sequence of daily-updated magnetograms as a background into which high-pressure structures without any internal magnetic field are added to resemble CME-related solar-wind disturbances. The reliability of ensemble CME-arrival predictions is strongly conditioned by the initial CME input parameters, such as width, direction, and speed \citep{Mays2015,Kay2020}, but also by the errors that can appear in the ambient-model parameters, and by the uncertainty of the solar-wind background derived from coronal maps. To improve the characterization of the heliosphere, multipoint coronagraph observations are used to infer CME parameters, using the GCS model, as described in Section \ref{sec:RS_Recons}. The inner boundary condition is given by the WSA V5.2 model, which uses standard quick-reduce zero-point corrected magnetograms from the Global Oscillations Network Group \citep[GONG,][]{Harvey1996GONG}, available from the National Solar Observatory website\footnote{\url{ftp://gong2.nso.edu/QR/zqs/}}. 

The ENLIL simulation interval encompasses about two days before the onset of CME1, to consider the preconditioning of the heliosphere and the prior IP structures that might be present in the heliosphere, and about ten days after the CME1 onset to follow the evolution of ICME1 and ICME2 in the IP medium. Table \ref{tab:GCSENLIL} summarizes the CME parameters obtained from the GCS reconstruction (columns 2--6 and 10) and the translation to ENLIL input (remaining columns). Values resulting from the GCS reconstruction for CME 01 and CME 02 (L. Balmaceda, private communication) are also listed in Table \ref{tab:GCSENLIL}, since they are taken into account for the ENLIL simulations to set, as mentioned above, more realistic solar wind conditions for CME1 and CME2. We note that the CME LE speed is chosen instead of the bulk speed (bright core, if present), as it often better captures the global and strong impact of the high-pressure structures. Regarding CME2, the final GCS reconstruction parameters were chosen based on a better reconciliation with in-situ data (see Section \ref{sec:RS_Recons}), as observed by PSP and STEREO-A (see Figure \ref{Fig:ENLIL_arrival} described below). The ENLIL simulation input parameters and results are available on the Community Coordinated Modeling Center (CCMC) website\footnote{\url{https://ccmc.gsfc.nasa.gov/database_SH/Laura_Rodriguez-Garcia_051121_SH_1.php}}. 

Figure \ref{Fig:ENLIL_arrival}a,b shows the comparison between the in-situ magnetic field data and over-plotted in pink line the result of the ENLIL simulation at the locations of PSP (Figure \ref{Fig:ENLIL_arrival}a) and STEREO-A (Figure \ref{Fig:ENLIL_arrival}b). From the top, both plots include the magnetic field strength, the magnetic field latitudinal and azimuthal angles, $\theta$\textsubscript{B-RTN} and $\phi$\textsubscript{B-RTN}, solar wind proton speed, temperature and density. The one-minute-averaged magnetic field observations are from FIELDS instrument onboard PSP (left) and from the Magnetic Field Experiment and plasma experiment onboard STEREO-A (right). At the location of STEREO-A, Figure \ref{Fig:ENLIL_arrival}b shows that ENLIL follows the general trend and mean magnitude of the observed magnetic field strength (first panel), solar wind speed (fourth panel) and temperature (fifth panel). The IP ICME2-driven shock arrival, indicated by the vertical blue line, is simulated to be a few hours late, but within the uncertainty of the model of the mean absolute arrival time prediction error is 10.4$\pm$0.9 hr \citep{Wold2018}. The changes in magnetic field polarity are reasonably well simulated (third panel). In the case of PSP, Figure \ref{Fig:ENLIL_arrival}a shows that the ICME1-driven shock arrival is simulated a few hours earlier, but again within the mean uncertainty, and the simulated magnetic field strength is similar in magnitude to the in-situ observations. This is not the case for the second structure observed by PSP, ICME2. ENLIL is simulating the shock arrival within the given accuracy, but the simulated magnetic field strength is much lower. \textbf{The interaction between ICME1 and ICME2 at the location of PSP could be behind this behavior and the lack of internal magnetic field for the CME driver}. Figure~ \ref{Fig:ENLIL_arrival}c shows a selected time frame from the ENLIL simulation, at the time of the ICME2-driven shock arrival at the location of PSP. The ecliptic plane view shows how the shock driven by ICME2 might interact with the trailing edge of ICME1 at the location of PSP. The meridional plane view presents the ICME1 and ICME2 interaction near the location of PSP. Under this simulation, the interaction is happening in a period ranged between November 30 $\sim$12 UT, as seen on the southern plane (based on the visual inspection of the simulation), and November 30 $\sim$23 UT, as seen from the ecliptic plane. 

The second model used for the prediction of the IP CME travel and its arrival at an arbitrary ecliptic-plane location is the DBM tool\footnote{\url{http://swe.ssa.esa.int/web/guest/graz-dbm-federated}}. The DBM uses two user-defined parameters, the solar wind speed and the drag parameter ($\Gamma$), which defines the velocity-change rate. The average solar wind speed was taken from ENLIL simulation and the $\Gamma$ parameter is 0.1 and 0.15 for CME1 and CME2, respectively \citep[see more details for the $\Gamma$ tuning in][]{Dumbovic2019}. The bulk speed has been considered as the true speed of the CME to infer the ICME arrival to the different spacecraft. The bulk speed represents the variation of OC\textsubscript{1} distance in time \citep[see Figure 1 in][]{Thernisien2011}, so that it corresponds to the true speed of the flux-rope structure. Figure \ref{fig:DBM} shows the CME radial distance  and speed of CME1 (pink) and CME2 (brown) based on DBM equations \citep[][]{Dumbovic2021}. The trailing edge of ICME1, indicated with the dashed pink line, is crossing the bulk of ICME2 represented by the brown dash-pointed line. This means that the trailing edge of ICME1 is colliding with ICME2 near PSP at ${\sim}$02:20 UT location (0.81 AU), indicated by the horizontal gray dashed line. The time of the interaction is highlighted on the figure and coincides with the arrival of the ICME2 (second red vertical line). The bottom panel in Figure \ref{fig:DBM} shows that the average speeds of ICME1 and ICME2 are 400 km s\textsuperscript{-1} and 800 km s\textsuperscript{-1}, respectively. These values will be the input speed for the in-situ reconstruction in next section.

\subsection{Modeling Based on In-Situ Observations and 3D Reconstruction}

For the reconstruction of the events based on in-situ observations, we have selected two versions of the same model and reconstruction techniques. The goal is to compare which model may reproduce better the magnetic field profile. The Elliptical-Cylindrical (EC) model \citep{Nieves-Chinchilla2018ApJ} is an upgraded version of the Circular-Cylindrical (CC) model \citep{Nieves-Chinchilla2016} that assumes axial symmetry, but allows distortion in the flux-rope cross-section \textbf{that is consistent with kinematic analysis of the CME evolution \citep[][]{Owens2006,Owens2006model}. The distortion or deformation of the CMEs in the heliosphere is an open question that is not solved yet, we have therefore done the comparison of the two models. The impact of the distortion assumption in the increase of the number of parameters is also a topic of debate. This is, in part, something addressed here: is the asymmetric magnetic field profile in the ICME2 due to distortion, expansion, or both?}

The model equations are, 
\begin{eqnarray}
\label{Eq:ECmodel}
    B_r&=& 0\nonumber\\
    B_{y}&=& \delta B_y^0\Bigg[\tau - \bigg(\frac{r}{R}\bigg)^2\Bigg] \\
    B_{\varphi}&=& -H \frac{2\delta h}{\delta^2+1}
     \frac{B_y^0}{|C_{10}|}\frac{r}{R} \nonumber
\end{eqnarray}
where \textbf{$R$ is the cross-sectional radius obtained from the reconstruction. $\delta$ is the distortion, $\delta B_y^0 \tau$ is the central magnetic field, $C_{10}$ in-situ the ratio between the poloidal and axial current density, $h=[\delta^2\sin^2\varphi + \cos^2\varphi]^{1/2}$ and $H$ is the chirality.}

This model does not enforce any force-free condition for the internal dynamics of the structure, but it provides information about the internal Lorentz force distribution ($C_{10}$) and the central magnetic field can be evaluated ($B_y^0$). The $H$ is the flux-rope chirality. The cross-section distortion is described with the $\delta$ parameter and the $\tau$ parameter provides information about the internal twist distribution. This last parameter, together with $C_{10}$ determines the stability of the structure \citep[see ][]{Florido2020}. According to the results of this analysis, we have fixed $\tau=1.1$.  

This reconstruction was first introduced by \citet{Hidalgo2000} and more details can be found in \citet[][]{Nieves-Chinchilla2016,Nieves-Chinchilla2018ApJ}. The reconstruction technique enables inference of the trajectory of the spacecraft while crossing the structure. From this procedure, we can obtain the axis azimuth, tilt and rotation around the axis ($\phi,\theta,\xi$), as well as the closest approach to the axis ($y_0$). For the reconstruction of the magnetic obstacle of ICME2, we have considered the solar wind bulk speed obtained from the DBM model in section~\ref{Sec:enlil}. 

\begin{figure}[ht!]
    \centering
    \includegraphics[width=0.7\textwidth, angle=90 ]{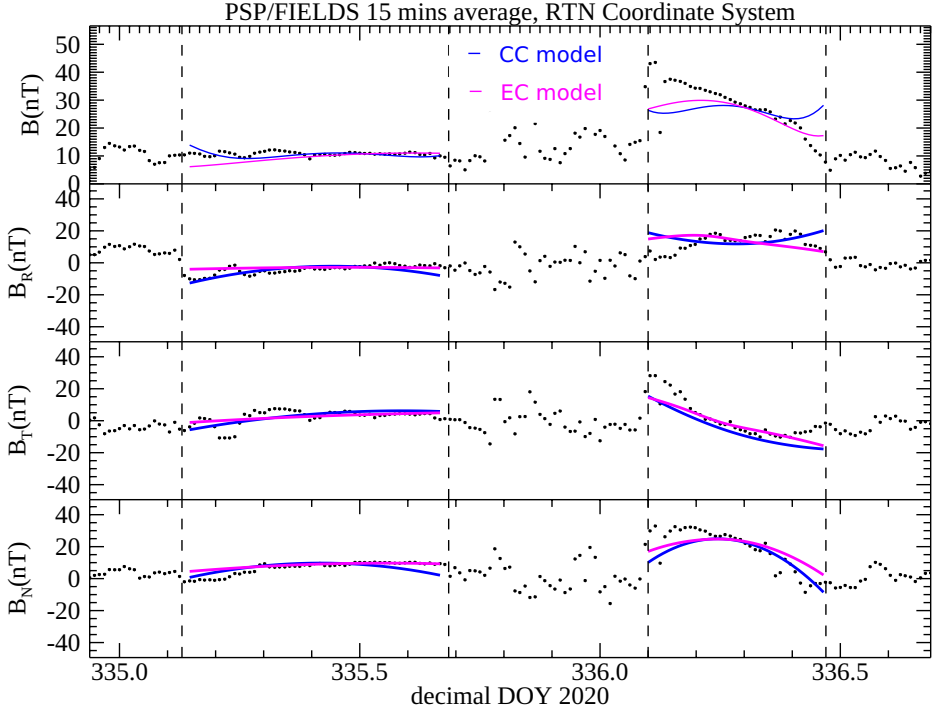}
        \caption{(a) Comparison of in-situ fitting results from the CC model (blue) and EC model (pink) with PSP magnetic field observations spanning the two events. From the top, the panels display the magnetic field strength and the three magnetic field RTN components, respectively.}
    \label{fig:ISfit}
\end{figure}

The reconstruction is initialized following the steps indicated in \citet{Nieves-Chinchilla2016} for the CC model and using the output parameters as the input parameters for the EC model. Figure~\ref{fig:ISfit}a 
compares the magnetic field magnitude and components observed by PSP and obtained from the output fitting parameters included in Table~\ref{Tab:Parmdl} to visually evaluate the goodness of fit. The top panel in the top figure is the magnetic field strength, followed by the three RTN magnetic components in separated layers, for the CC model in blue and for the EC model in pink. Despite the fact that the first event does not display significant signatures of distortion, the EC model captures the trend of the magnetic field components, \textbf{as expected}. In the case of ICME2, the obvious asymmetry in the magnetic field strength cannot be captured by the CC model, but the EC model follows better the magnetic field strength as well as its components. Table~\ref{Tab:Parmdl} lists the output parameters obtained from the reconstruction. From column $H$ to column $y_0/R$, the table includes the direct parameters obtained from the reconstruction technique using the CC and EC model in both events. \textbf{The angles are in RTN coordinate system.} It is followed by the size of the cross-section, $R$. \textbf{The cross-section radius is obtained from the reconstruction. Once the trajectory of the spacecraft is inferred, knowing the duration and assuming constant bulk velocity, this value can be obtained.  The last two columns include the goodness parameters, $\chi^2$ and $\rho$ \citep[see][for details]{Nieves-Chinchilla2018SolPhy}, which are consistent with our visual inspection analysis. In general, the EC model fits better the magnetic field profile for both ICMEs.}

\begin{figure}[ht!]
    \centering
    \includegraphics[width=0.95\textwidth]{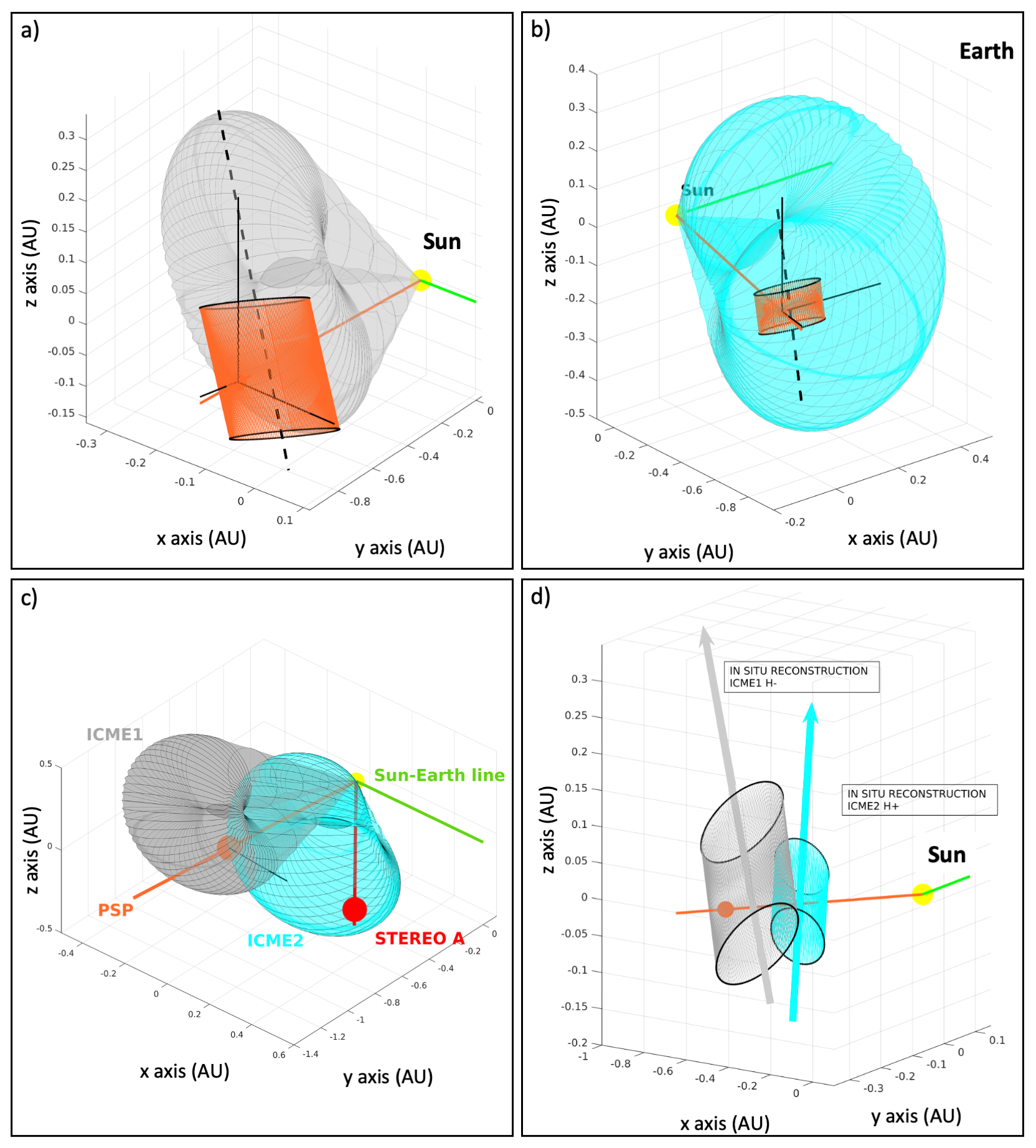}\\
        \caption{
        (a) Comparative view of in-situ (orange) and remote-sensing (gray) reconstructions of ICME1. (b) Comparative view of in-situ (orange) and remote-sensing (cyan) reconstructions of ICME2. Both (a) and (b) have remote-sensing reconstructions assuming self-similar expansion from the lower Corona until they reached PSP (orange). (c) ICME1 on the left (gray) and ICME2 on the right (cyan). General view of the reconstructions of the interaction between ICME1 and ICME2 at the PSP position. We have expanded the ICMEs to this height based on the latest GCS reconstruction and the inferred speed from Table~\ref{Tab:Parmdl}. (d) General view of the reconstruction of the interaction of the ICME1 (gray) and ICME2 (cyan) at the PSP position. Red and orange lines show the crossing trajectory of ICME1 and ICME2 with PSP and STEREO-A, respectively.}
        
    \label{fig:reconstruction}
\end{figure}

In terms of the orientation and geometry of the reconstruction, we have plotted the 3D reconstruction based on in-situ and remote-sensing observations in Figure~\ref{fig:reconstruction} to facilitate the interpretation by comparing GCS with the EC model. The figure displays the two reconstructions merged in the same graphic with the remote-sensing reconstruction self-similarly propagated to the PSP height.  Figure~\ref{fig:reconstruction}a shows the combined in-situ reconstruction and the gray torus for the remote-sensing reconstruction of CME1/ICME1 and shows good agreement between the reconstructions.  Figure~\ref{fig:reconstruction}b encloses the in-situ 3D reconstruction in orange, and the 3D remote-sensing reconstruction in cyan for ICME2. In contrast to the reconstruction of ICME1, the ICME2 reconstruction does not agree so well. Among possible causes of this discrepancy may be underestimation of the bulk velocity, the fact that the model is static, but mainly that the remote-sensing 3D reconstruction does not reflect the actual 3D shape of the CME. Finally, Figure~\ref{fig:reconstruction} shows the remote-sensing (panel c) and in-situ (panel d) reconstructions for both ICME1 and ICME2 at the time when ICME2 reaches the PSP location. We have expanded the CMEs to this height based on the latest GCS reconstruction and the speed listed from Table~\ref{Tab:Parmdl}. It is possible to see that, due to the interactions of both ICMEs at the PSP location, in-situ and remote-sensing observations are impacted. This interaction leaves a more complex structure in the measurements, which requires more complex modeling tools.

\begin{table}
\centering
\scriptsize

 \begin{tabular}{|l ccc |lcccccccc| c |cc|}
 \hline
 Event - mdl & $v_{sw}$ & Start  & Dur & H & $C_{10}$  & B$_{c}$ & $\tau $& $\delta$ & $\phi$  & $\theta$ & $\xi$ &  y$_{0}$/R &  R & $\chi^{2}$  & $\rho$  \\
 &  km s$^{-1}$& dDOY & hrs & & & (nT) & & &($^{o}$)  & ($^{o}$) & ($^{o}$) &   &  (au)  &   &  \\
 \hline 
ICME1   $CC$    & 400 & 335.13  & 13.32  & $-$ & 4.70 & 47 & 1.1 & 1. & 28 &  42 & $-$ & 0.92 & 0.0893 & 0.43 & 0.38\\
$\;\;\;\;\;\;\;\;\;\;\;\; EC$    &   &  &  &  $-$ & 5.90 & 36 & 1.1 & 0.64 & 61 &  66 &  52 & 0.88 & 0.249 & 0.40 & 0.52\\
ICME2   $CC$    & 800 & 336.10  & 8.88  & $+$    &  1.78  & 47  & 1.1& 1. & 217 &  67 & $-$ & 0.75 & 0.1095 & 0.32 & 0.33\\
$\;\;\;\;\;\;\;\;\;\;\;\; EC$    &   &  &  &  $+$ & 1.48 & 32 & 1.1 & 0.58  & 13 &  76 &  11 & 0.38 & 0.1153 & 0.28 & 0.61\\
\hline
\end{tabular}
\smallskip
 \caption{Output parameters from the in-situ reconstruction technique. The table includes information used for the reconstruction, the set of output parameters in RTN coordinate system, and the goodness of the fitting procedure. From the left: the event, model, average bulk solar wind speed ($v_{sw}$), start time (dDOY) in decimal DOY, duration ($Dur$) in hours, $H$, $C_{10}$, central magnetic field ($B_{c}$), $\tau$ and distortion ($\delta$); axis azimuth ($\phi$), axis tilt ($\theta$), rotation around the axis ($\xi$), the closest distance to the axis relative to the flux rope radius, $y_{0}/R$, cross-sectional radius (R) and the goodness parameters, $\chi^{2}$ and $\rho$. }
\label{Tab:Parmdl}
\end{table}

\begin{figure}[ht!]
    \centering
    \includegraphics[width=0.5\textwidth]{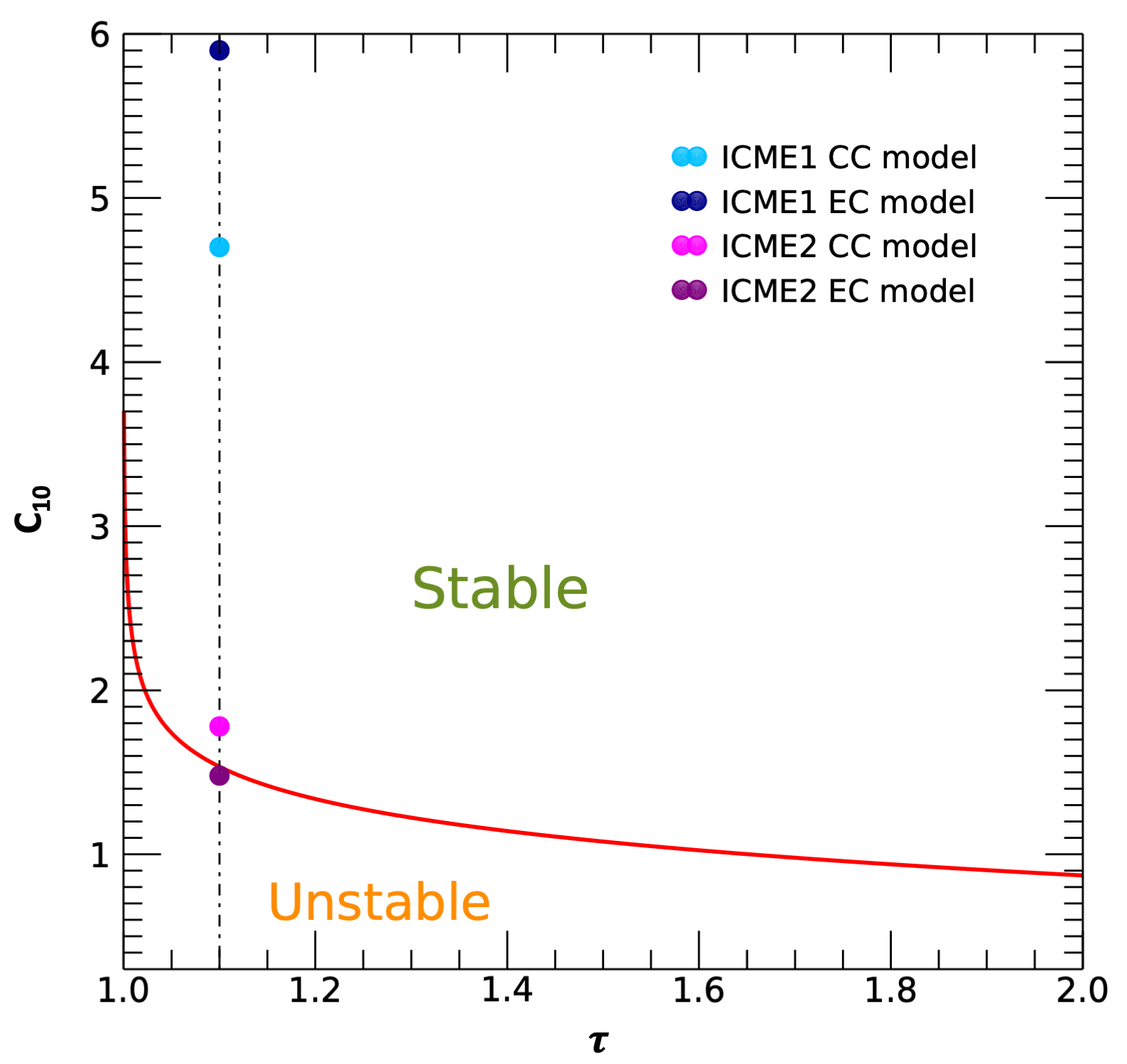}
        \caption{Flux rope kink stability based on the range of values of $C_{10}$ and $\tau$. For $C_{10}-\tau$ values above the red line, the structure remains stable. For values below the red line, the structure is kink unstable. The reconstructions included in this paper are constrained to $\tau=1.1$ (vertical dashed line). The colored dots over the dashed line are the values obtained from the reconstruction of both ICMEs and based on both CC and EC models.
        }
    \label{fig:stability}
\end{figure}

Finally, we evaluated the flux rope kink-stability that may cause rotation of the flux ropes at the PSP heliocentric distance. Based on Equation~12 in \citet{Florido2020}, and the model equations \ref{Eq:ECmodel}, the threshold to assume a stable structure kink follows, 

\begin{equation}
    C_{10}=\frac{5.6382*0.84}{5.7}
    \bigg( \frac{\tau-1}{5.7}\bigg)
    e^{-(\frac{\tau-1}{5.7})^{0.84}} ,
\end{equation}

where $\tau=1.1$ in our reconstructions. Figure~\ref{fig:stability} displays the threshold between the stable and unstable regimes for the pair of $C_{10}$ and $\tau$ values for the CC model \citep[see for more details][]{Florido2020}. For $C_{10}-\tau$ values above the red line, the structure remains kink-stable. For values below the red line, the structure is kink-unstable and therefore carries out deviations from the radial propagation, specifically rotations. The colored dots over the dashed line are the obtained $C_{10}$ values of the reconstructions of both ICMEs, based on both CC and EC models and included in Table~\ref{Tab:Parmdl}. All $C_{10}$ values fall within the kink-stable range except the value for the ICME2 reconstruction using the EC model. This is a marginal value of $C_{10}=1.48$, where the threshold for $\tau=1.1$ is $C_{10}$=1.53. This result suggests that the CME1--CME2 interaction may be driving a rotation in ICME2 and, therefore, a deviation from a radial propagation.

\section{ICME1--ICME2 Interaction and Collision}
\label{Sec:discussion}

The analysis carried out and described in the previous section demonstrates that the interaction between both CMEs is in progress while PSP crosses both structures. Although ENLIL numerical modeling indicates that the collision started on November 30 at 12:00 UT out of the ecliptic plane, the DBM model indicated the collision started on December 01 at 02:20 UT. In this section, we  review and interpret the in-situ signatures according to the physical processes associated with the interaction and we also discuss an experiment to assess the effect of the expansion on the ICME2 magnetic configuration.

\subsection{In-Situ Signatures of the Interaction}

As of yet, there is not a significant amount of research related to the interaction or collision of ICMEs in the inner heliosphere based on remote-sensing observations \citep[e.g.][]{Lugaz2012b,Liu2014a,Colaninno2015,Kilpua2019,Palmerio2019}. Most of them are based on numerical simulations \citep[see][for a review]{Lugaz2017SoPh}. Again, this is in part due to the lack of observations of the events that limit the investigations. \citet{Gopalswamy2001ApJ} was one of the first studies that reported Type II radio observations of the interaction between CMEs and, simultaneously to the radio enhancement, a fast CME overtaking a slow CME when viewed in coronagraph images. Since then, type II radio observations have been regarded as one of the first signatures of CME--CME interaction \citep[see e.g.][]{Shanmugaraju2014,Temmer2014,Makela2016,Al-hamadani2017,Morosan2020} and have been used as a tool to explore the relationship with solar energetic particle acceleration and transport in the context of large-scale interactions \citep[e.g.][]{Gopalswamy2002,Li2012,Zhuang2020}.

While there are many physical processes associated with the interaction, such as particle acceleration or plasma heating, the collision among the magnetic obstacles is the peak of such encounter. In a simplistic view of two magnetized plasmoids colliding, the dynamic of such hit is basically determined by the kinematic aspects and the internal magnetic structure \citep[e.g.][]{Schmidt2004,Lugaz2005,Xiong2006,Shen2011,Shen2017,Niembro2019}.  Thus, the amount of momentum in each structure ($\Delta m\vec{v}$) plus the orientation and chirality of the structures in the solar wind would determine the consequences of the collision. Both CMEs will therefore co-exist and move together as they travel away from the Sun if there is a low momentum transfer and magnetic configuration such that magnetic reconnection is not allowed \textbf{(elastic collision)}. They will slowly merge together when such magnetic configuration is in opposite direction and enabling reconnection. A more dramatic scenario is when a fast and energetic CME overtakes a slow CME \citep[\textit{cannibalism},][]{Gopalswamy2001ApJ}. The trailing part of the front CME would be completely absorbed by the rear CME. Based on the following evidences, we believe the scenario \textbf{for the November 2020 event is that} of an elastic collision. This case could be described as a perfect elastic collision where the result of the momentum exchange may separate the two magnetized structures \citep[][]{Shen2012NatPh}. 

The results of our analysis indicate that PSP observed the immediate instants around the collision of the two CMEs. In our case, CME1 and CME2 are not a perfect magnetized ball, but two 3D structures with the second CME driving a shock wave. Here, CME2 is almost twice as fast as CME1 and, even though the size is not well characterized by our in-situ and remote-sensing reconstructions, the chirality and orientation is such that the magnetic field direction of the contact interface is the same (i.e., west in RTN coordinates) and therefore is not conducive to reconnection (see Figure~\ref{fig:reconstruction}). We conjecture that the transfer of momentum started before the CME--CME interaction, during the CME1--Shock2 interaction. 

In terms of the changes in the magnetic structures, we have identified two different features that would demonstrate the kind of physical processes acting on ICME1 and ICME2. ICME1 is overtaken by Shock2 that may have reached the structure faster than ICME2 because the lower density in the solar wind facilitates wave propagation \citep[see ][ and references within]{Lario2021}. This interaction may accelerate the leading structure and momentarily increase the distance with ICME2. Thus, this may explain the depleted magnetic field strength in the sheath region between Shock2 and MO2. This scenario would confirm that PSP observed the instants immediately preceding the collision between the magnetic structures. The magnetic field components profile suggest that Shock2 had already initiated the penetration inside ICME1 and would eventually reach Shock1. This is one of the scenarios described by \citet[][]{Shen2011} of the three-dimensional time-dependent, numerical MHD simulations of the interactions of two ICMEs. Plasma measurements taken by PSP, albeit sporadic and almost entirely missing after the passage of Shock2 (see Figure~\ref{fig:IS}), can help \textbf{shed} additional light on the complex processes at interplay. First, the speed profile of MO1 (bottom-most panel in Figure~\ref{fig:IS}) does not show signatures of expansion with the exception of a weakly decreasing trend during the first third of the structure. This suggests that the approaching Shock2 may have halted the expansion of ICME1 as it started its penetration. Secondly, the region immediately following the trailing edge of MO1 (i.e., the Shock2 upstream) sees an increase in temperature with respect to the preceding period. Although this may simply indicate that PSP has left the core of ICME1, plasma heating is a known effect of shock--CME interactions in the solar wind \citep[e.g.,][]{Vandas1997,Farrugia2004}. Given the overall magnetic field profile following the trend of MO1 mentioned in Section~\ref{subsec:insitu}, it is possible that the Shock2 upstream marks an interaction region that is in the process of being disturbed by the passage of the following ICME.

In the case of ICME2, there are two significant features, i.e.\ the drop at the front of the magnetic structure for $\sim$15 minutes and the asymmetric profile in the magnetic \textbf{field strength}. According to \citet[][]{Feng2013}, the first of these signatures can be associated with magnetic reconnection exhausts within the interior of ICMEs \citep[see e.g.][]{Xu2011}.  \textbf{The} ongoing magnetic reconnection may cause heating in the internal plasma of ICME2 and drive CME over-expansion \citep[][]{Odstrcil2003,Schmidt2004,Chatterjee2013,Lugaz2013}.

As stated in \citet{Lario2021}, this event features significant similarities with the Bastille Day event at Earth \citep[see e.g.][]{Lepping2001,Smith2001} in both magnetic field and energetic particle signatures. Thus, we have inspected the available data and compared both cases. Magnetic field strength and density depletion with an increase in the proton temperature in the sheath region of the rear, faster ICME supports the scenario described above. The solar wind bulk velocity within the second magnetic structure indicates significant expansion, with a value of ${\Delta}V \sim 180$~km~s$^{-1}$. The speed profile within the first ejecta, on the other hand, shows weak signatures of expansion, suggesting that the following shock had not yet started to profoundly affect its propagation.

\subsection{ICME2 Over-Expansion Assessment}

Visual inspection of both structures indicates that, while the magnetic field strength configuration of ICME1 does not display signatures of distortion with a very flat and symmetric magnetic field strength, ICME2 displays a very asymmetric profile. As stated in \citet{Nieves-Chinchilla2018SolPhy}, there are three main causes for asymmetric profiles in ICMEs: distortion, erosion, or expansion. In the case of distortion, the spacecraft could enter the structure by the squeezed region but exit by the elongated region. In the case of erosion, part of the magnetic structure has been \textbf{peeled away via} magnetic reconnection with the ambient solar wind \citep[][]{Dasso2005,Ruffenach2012JGR}.  Finally, in the case of significant expansion, \textbf{the aging} \citep[decrease in the magnetic field strength during the spacecraft transit, see][among others ]{Farrugia1992,Osherovich1993,Demoulin2020} should be taken into account in the modeling but also the effect of expansion on the spacecraft measurements and therefore on the reconstruction.  Here, one may think in an analogy with the Doppler effect, as expansion velocity positively contributes to the ICME bulk velocity, it has an artificial effect of magnetic field strength compression in the in-situ observations. Conversely, when the expansion velocity component is opposite to the structure bulk velocity, the effect is of dilatation and decrease in the measurements.  As previously stated, the lack of plasma observations prevents us from concluding which, among the three causes, may be responsible and subsequently reconstruct the structure.  The implications for the reconstruction of the structure will be discussed here.  

In general, ICME expansion can be defined using the difference in velocity between the ICME front and the rear.  The cases with over-expansion \textbf{have expansion velocity above an average of} 25 km s$^{-1}$ at 1 AU \citep[e.g.][]{Owens2005,Nieves-Chinchilla2018SolPhy,Gopalswamyetal2015JGRA}. 
Over-expansion, first observed by \citet[][]{Gosling1994,Gosling1998}, has been identified as one of the secondary effects on the CME--CME collision \citep[][]{Schmidt2004,Lugaz2013}. The cause of the increase of the internal plasma pressure could be related to the energy released by magnetic reconnection processes \citep[][]{Murphy2011}. 

The in-situ reconstruction based on cylindrical geometry with circular and elliptical cross-section seemed to be insufficient to fully describe the magnetic configuration. Thus, the experiment we describe here aims to evaluate, qualitatively and quantitatively, the impact of expansion on the CME magnetic configuration. Thus, assuming magnetic flux and helicity conservation, the aging of the magnetic configuration can be quantified as:

\begin{eqnarray}
\label{Eq:Eqexp}
    B_y{'}&=\bigg(\frac{R}{R{'}}\bigg)^2B_y,\nonumber \\
    B_\varphi{'}&=\frac{R}{R{'}}\frac{L}{L{'}}B_\varphi, 
\end{eqnarray}

\citet{Florido2020} also demonstrated the effect of the aging on model parameters as,

\begin{eqnarray}
\label{Eq:Parexp}
    B_y^0{'}&=&\frac{R}{R'} B_y^0, \nonumber \\
    \tau{'}&=&\tau ,\\
C{'}_{10}&=&\frac{L{'}}{L}\frac{R}{R{'}}C_{10}.\nonumber
\end{eqnarray}

Here $L$ and $L^{'}$ are the non-expanded and expanded length of the flux rope axis. The $R$ and $R^{'}$ are non-expanded and expanded cross-section radius. For both quantities, the rate of change is $R^{'}=R+V_\mathrm{exp}^{R} t_s$ and $L^{'}=L+V_\mathrm{exp}^{L} t_s$, where $t_s$ is the spacecraft transit time through the structure. The velocity of expansion of the radius ($V_\mathrm{exp}^{R}$) and flux-rope length ($V_\mathrm{exp}^{L}$) is generally assumed to be equal and constant when it is assumed there is a radial propagation and self-similar expansion of the structure in the solar wind. 

Assuming self-similar expansion ($L/L^{'}=R/R^{'}$), the above set of equations~(\ref{Eq:Eqexp}) and~(\ref{Eq:Parexp}), reduce to,

\begin{eqnarray}
\label{Eq:Eqexp1}
    B_y{'}&=&\bigg(\frac{R}{R{'}}\bigg)^2B_y,\nonumber \\
    B_\varphi{'}&=&\bigg(\frac{R}{R{'}}\bigg)^2 B_\varphi, \text{ and } \nonumber \\
    B_y^0{'}&=&\frac{R}{R'} B_y^0, \\
    \tau{'}&=&\tau ,\nonumber \\
    C{'}_{10}&=&C_{10}. \nonumber
\end{eqnarray}
Note that the model parameter $C_{10}$ remains constant if the self-similar condition is assumed.

\begin{figure}[ht!]
    \centering
    \includegraphics[width=0.5\textwidth]{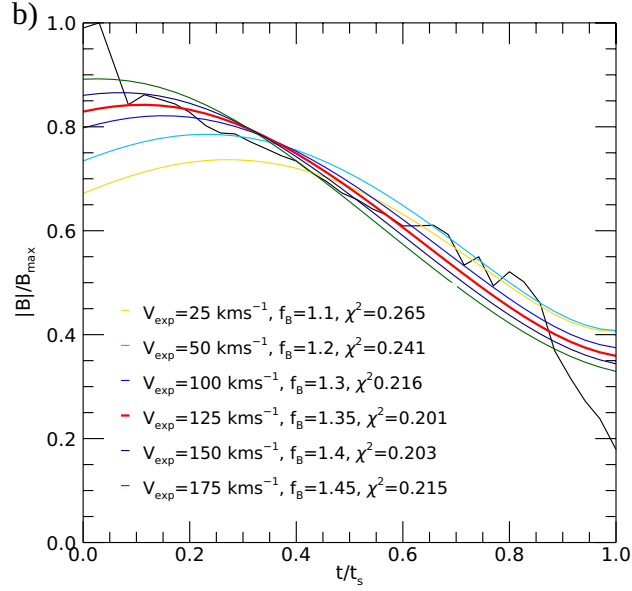}
    \caption{Plot of ICME2 magnetic field strength with overplots of the magnetic field strength based on different $V_\mathrm{exp}$ and $f_B$ obtained from the spacecraft trajectory simulation crossing the flux rope. }
    \label{fig:exprofile} 
\end{figure}

To evaluate the impact of the expansion on the magnetic field configuration, we have simulated the trajectory of the spacecraft and reproduced the change in the magnetic field configuration under different $V_\mathrm{exp}$ values. The reconstruction technique also requires taking into account the effect of the expansion in the spacecraft observations. The change in size of the cross-section ($R^{'}=R+v_{sw}t$) at each instant the spacecraft is within the flux rope, is affected by the position of the spacecraft at the entrance of the flux-rope, which, in turn, is affected by the expansion.  Thus,

\begin{equation}
\label{Eq:kk2}
   x_0 = \frac{(v_\mathrm{sw}-V_\mathrm{exp})}{2}t -
   \frac{[a\sin\phi\sin\theta\cos\theta]+
      (\delta^2-1)\cos\phi\cos\theta\sin\xi\cos\xi]}
      {(a\sin^2\phi\sin^2\theta +b\cos^2\phi)
      +2(\delta^2-1)\sin\phi\cos\phi\sin\theta\sin\xi\cos\xi}z_0,
\end{equation}
where, $t$ is the time at which the spacecraft is within the structure and 

\begin{eqnarray}
    a=[\delta^2\cos^2\xi+\sin^2\xi] \nonumber \\
    b=[\delta^2\sin^2\xi + \cos^2\xi].
\end{eqnarray}
Note that, as the $V_\mathrm{exp}$ value increases, the effect of aging of the structure is also more significant and therefore a scale factor ($f_B$) on the $B_c$ should also be quantified. Figure~\ref{fig:exprofile} illustrates the analysis in the case of the EC model. For each $V_\mathrm{exp}$, we identified a $f_B$ that minimizes the equation,

\begin{equation}
    \chi^2=\frac{\sum (\sqrt{B^2_\mathrm{obs}-B^2_\mathrm{sim}})}{N},
\label{Eq:chi}
\end{equation}
where $B^2_\mathrm{obs}$ is the magnetic field strength observations, $B^2_\mathrm{sim}$ is the simulated observations, and $N$ is the number of points included in the analysis. 
For each ($V_\mathrm{exp}, f_B$) pair, we identified the pair that minimizes the equation~(\ref{Eq:chi}). Figure~\ref{fig:exprofile} illustrates this. Each colored line over plotted on the magnetic field observations represents the pair $V_\mathrm{exp}, f_B$, which minimizes equation~(\ref{Eq:chi}) for such $V_\mathrm{exp}$. The red color line is the one that better reproduces the magnetic field strength profile \textbf{($\chi^{2}=0.201$)}. Therefore, the results demonstrate that, for the EC model, \textbf{the velocity of} expansion required to fully reproduce the magnetic field strength is $V_\mathrm{exp}=125$ $km~s^{-1}$. Additionally, the central magnetic field strength ($B_c$) included in Table~\ref{Tab:Parmdl} should be multiplied by the factor $f_B=1.35$. The resulting value is $B_C^\mathrm{exp}=43.2$ nT.

\begin{figure}[ht!]
    \centering
    \includegraphics[width=0.9\textwidth]{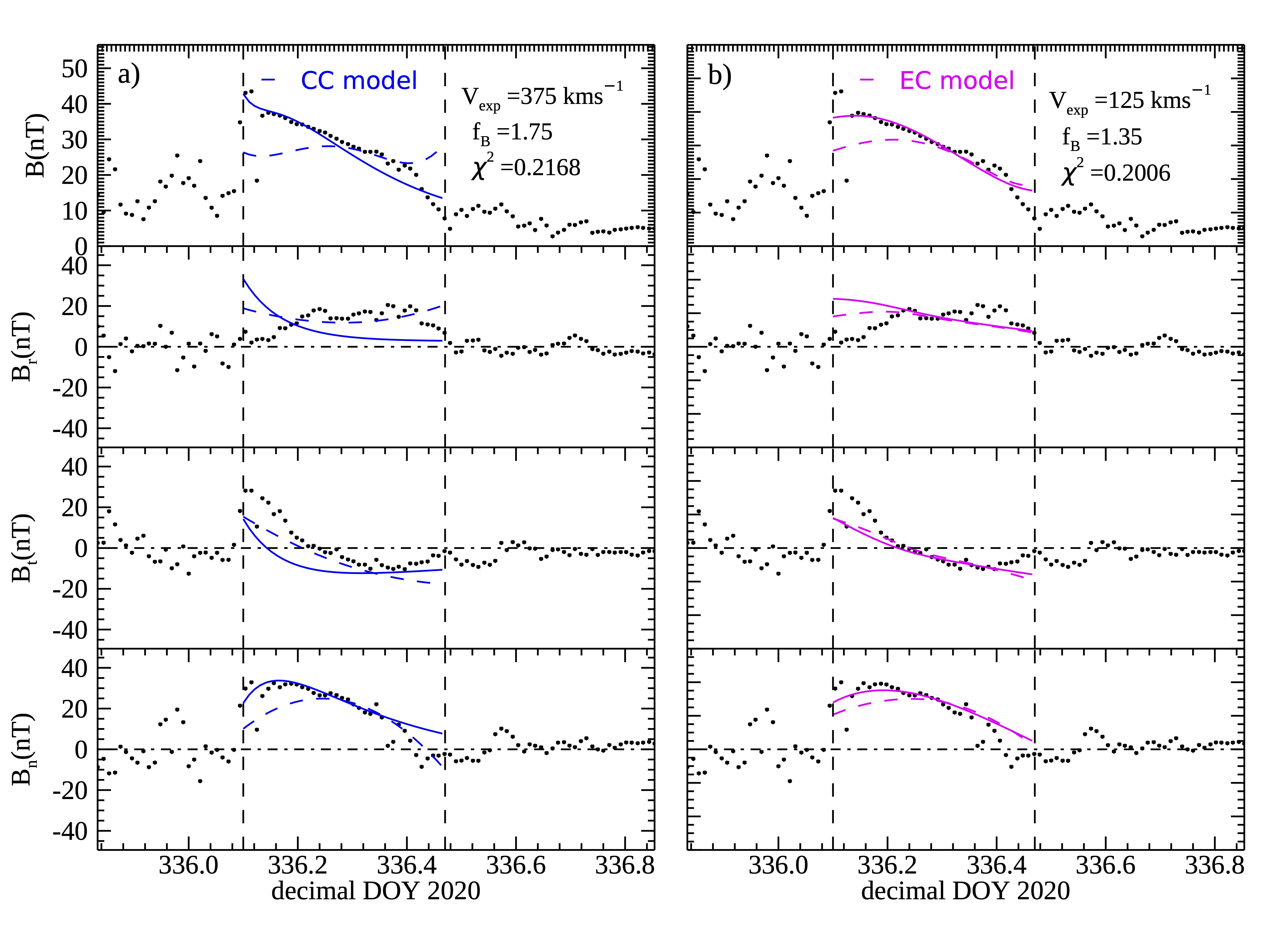}
    \caption{Comparative analysis of the effect of expansion on the magnetic field configuration of ICME2. (a) Based on the CC model, and (b) based on the EC model. Over plotted on the ICME2 magnetic field strength and components, the dashed colored lines depict the CC (blue) and EC (pink) model without expansion. The over plotted colored lines illustrate the effect of the expansion on the magnetic field configuration of the magnitude and components. }
    \label{fig:CompExp}
\end{figure}

Similar exercises have been done with the CC model. Figure~\ref{fig:CompExp} shows a comparative analysis of the effect of the expansion on the magnetic field magnitude and components for the CC model (Figure~\ref{fig:CompExp}a) and EC model (Figure~\ref{fig:CompExp}b). The colored lines display the CC model (blue) and the EC model (pink) with expansion and the dashed lines represent the models without expansion.  In the CC model case, the expansion velocity needed to reproduce the magnetic field strength profile is $V_\mathrm{exp}=375$ km~s$^{-1}$ with a $f_B=1.75$. The maximum expansion velocity found by \citet{Nieves-Chinchilla2018SolPhy} in the ICMEs observed by the Wind spacecraft during 1994--2015 was 270 km~s$^{-1}$. Thus, this value is significantly higher than the expected values for an expanding flux rope at 1~AU. The expansion itself is not sufficient to describe the ICME2 internal structure and we can conclude that the addition of expansion to a distorted geometry is needed.

\section{Summary, Conclusions and Final Remarks}
\label{Sec:conclusions}

The scarcity of multipoint and multiview observations of CMEs in the heliosphere leads us to exploit each observation opportunity and to challenge our current understanding of these large scale structures in the solar wind. We lack an understanding of the physical characteristics of the internal structure of CMEs,  how they are connected with the formation processes or how the innate CME features connect with the matured CME features in the heliosphere. 
The period of 2020 November 24-29 was a period of multipoint and multiview opportunities that enabled us to address these questions in relation to the CMEs observed. During this period, PSP was located at 104$^{\circ}$ west and at 0.81 AU and STEREO-A at 65$^{\circ}$ west as seen from Earth. Telescopes onboard STEREO-A and Earth-based observatories identified successive expulsions of CMEs that drove the first widespread solar energetic particle event of Solar Cycle 25. \textbf{Four of these CMEs were in the PSP direction, but just two of them impacted the spacecraft directly and are the focus of this study. However, only the flank of the second one skimmed STEREO-A}.

In this paper, we have assembled observations, models and techniques in a campaign-event exercise to understand the portion of the life-cycle of the two ICMEs (ICME1 and ICME2) up until they reach the PSP spacecraft \textbf{from November 29 to December 1}. The results of our analysis concluded that the two CMEs were under an interaction process during their passage by PSP. This conclusion is driven by the analysis of remote-sensing and in-situ reconstructions, combined with CME propagation modeling. Several iterations were carried out to reconcile these three elements of analysis during which two interaction times were identified based on the WSA-ENLIL+Cone model and DBM model that concurred to within a range of hours. Based on the in-situ reconstruction, and supported by the fact that the speed of ICME2 is almost twice as much as that of ICME1, we have conjectured that the two structures of different polarity and similar orientation \textbf{would result from an elastic collision}. The result of this collision does not appear to conduct any cannibalism but may result in a momentum exchange with a separation of the two MOs. 

The above conclusion is also supported by our interpretation of the observations. In terms of the radio emissions, this event was widely observed by all heliospheric observatories and the associated type II burst exhibited significant left-hand circular (LHC) polarization, \textbf{which has never been} detected before in this frequency range. The complex type III observations were followed by an intermittent type II burst that suggested an interaction of the CME2-driven shock with the base of the legs of CME1. We also analyzed frequency drift and in the case of Wind, this indicates that the PSP arrival time is also within the range calculated time by CME propagation models. 

In terms of the magnetic field and plasma parameters observed for ICME1, the analysis of the leading shock of ICME1 (shock1) indicates that it is still in the early \textbf{stages of formation}. The MO1 speed profile does not show signatures of expansion, 
but a small jump \textbf{at $\sim$07:30 on November 30, however},suggests that the approaching shock2 may have halted the expansion. In the interval around shock2 and the preceding area, there is an increase in plasma temperature but the magnetic field configuration appears to follow the trend of the MO1 magnetic field configuration. It also suggests that shock2 is heating the plasma within the MO1 and perturbing the magnetic field of MO1. In the case of ICME2, the depletion in the sheath magnetic field may be due to the prior passage of ICME1 as well as the separation between the two ICMEs. 

In terms of the magnetic field observations of ICME2, the internal structure also depicts signatures that may be associated with an interaction. We have analyzed the declining magnetic strength profile of MO2 and identified a clear signature of magnetic reconnection exhaust that, if accompanied by the plasma parameter, could confirm the early stages of an erosion process at the front. However, the steeped decline in the magnetic field strength profile cannot be explained by just the erosion. Thus, based on previous studies, we may conjecture that this profile represents over-expansion due to heat transfer from the reconnection process or to the distortion. Thus, we carried out an experiment to reproduce the asymmetry in the magnetic field of ICME2. We demonstrated that the elliptical cross-section distortion, or the expansion by themselves, are not sufficient to reproduce such a profile. \textbf{The method that reproduces the closest approximation includes both the distortion and the expansion}. For this case, we estimated an expansion of 125 km~s$^{-1}$, which is within the expected range of values for an over-expanding structure. Finally, in this paper, and for the first time, we have included an attempt to evaluate the change in the expected propagation based on the analysis of the internal properties of the CMEs. The marginal value of the $C_{10}$ parameter suggests that the MO2 may be in the range of instability and therefore suffering a deviation from the radial propagation. This is a very preliminary exercise that will require more detailed analysis in the future.

\begin{acknowledgments}
We are grateful for the dedicated efforts of the entire Parker Solar Probe team. We acknowledge NASA FIELDS contract NNN06AA01C, NASA SWEAP contract NNN06AA01C. The authors thank Dr. Antoinette Galvin of the University of New Hampshire, Principal Investigator of PLASTIC instrument on board STEREO-A and NASA contract NAS5-00132. T.J-Ch. acknowledges support from NASA under HGI Grant No. 80NSSC19K0274.
N.A. acknowledges support from NASA under HGI Grant No. 80NSSC20K1070.
H.C. is member of the Carrera del Investigador Cient\'ifico (CONICET) and acknowledges support from UTN projects UTI4915TC, UTN5445ME, and MSTCAME8181TC. L.F.G.S was supported by NASA Grant 80NSSC20K1580.
L.R.-G. is supported by the European Space Agency under the ESA/NPI program and acknowledges the additional financial support by the Spanish Ministerio de Ciencia, Innovación y Universidades FEDER/MCIU/AEI Projects ESP2017-88436-R and PID2019-104863RB-I00/AEI/10.13039/501100011033. T.N-Ch. and L.R.-G. also thanks Laura Balmaceda for her help.  L.F.G.S acknowledges support from NASA under Grant No. 80NSSC20K1580.
E.P. is supported by the NASA Living With a Star Jack Eddy Postdoctoral Fellowship Program, administered by UCAR’s Cooperative Programs for the Advancement of Earth System Science (CPAESS) under award no. NNX16AK22G. 
V.K. acknowledges the support by NASA under grants 18-2HSWO2182-0010 and 19-HSR-192-0143.
P.H., K.K.R., D.B.S, and T.N. acknowledge support from the NASA Heliophysics System Observatory Connect Program, grant 80NSSC20K1283.
ENLIL simulation results have been provided by the CCMC at NASA Goddard Space Flight Center (GSFC) through their public Runs on Request system (\url{http://ccmc.gsfc.nasa.gov}; run ID Laura\_Rodriguez-Garcia\_051121\_SH\_1). The WSA model was developed by N. Arge, currently at GSFC, and the ENLIL Model was developed by D. Odstrcil, currently at George Mason University.
\end{acknowledgments}

%

\vspace{5mm}
\facilities{PSP, SDO, SOHO, STEREO-A}




 
\newpage


\bibliographystyle{aasjournal}
\bibliography{manuscript_bib}{}



\end{document}